\documentclass[a4paper,11pt]{article}
\pdfoutput=1 

\usepackage{jheppub,bm} 

\usepackage[T1]{fontenc} 
\usepackage[utf8]{inputenc}

\newcommand{\slsh}{\rlap{$\;\!\!\not$}}     

\newcommand{\eps}{\epsilon}

\def\as{\alpha_s}

\def\beq{\begin{equation}}
\def\eeq{\end{equation}}
\def\beqn{\begin{eqnarray}}
\def\eeqn{\end{eqnarray}}

\def\hbbj{$H\rightarrow b\overline{b}j\;$}

\def\hbbqq{$H\rightarrow b\overline{b}q\overline{q}\;$}
\def\hbbgg{$H\rightarrow b\overline{b}gg\;$}
\def\hbbg{$H\rightarrow b\overline{b}g \;$}
\def\hbb{$H\rightarrow b\overline{b}\;$}

\def\yc{$y_{\rm{cut}}$\;}
\def\ycm{y_{\rm{cut}}}
\def\tauc{$\tau^{\rm{cut}}_3$\;}
\def\taucm{\tau^{\rm{cut}}_3}

\def\spa#1.#2{\langle #1 #2\rangle}
\def\spb#1.#2{[ #1 #2]}
\def\spab#1.#2.#3{\langle #1 |#2| #3] }

\allowdisplaybreaks[4]

\begin{document}

\title{$H \to b \overline{b} j$ at Next-to-Next-to-Leading Order Accuracy}

\author{Roberto Mondini}
\author{and Ciaran Williams}

\affiliation{Department of Physics,\\ University at Buffalo, The State University of New York, Buffalo
14260, USA}

\emailAdd{rmondini@buffalo.edu}
\emailAdd{ciaranwi@buffalo.edu}

\newcommand{\zero}{{(0)}}
\newcommand{\one}{{(1)}}
\newcommand{\two}{{(2)}}
\newcommand{\ztwo}{\zeta_2}
\newcommand{\zthree}{\zeta_3}
\newcommand{\cf}{C_F}
\newcommand{\ca}{C_A}
\newcommand{\nf}{n_f}
\newcommand{\cfs}{C_F^2}
\newcommand{\Tcm}{\tau_{cm}}
\newcommand{\TAij}{\mathbf{T}^a_{ij}}

\abstract{We present the calculation of the decay \hbbj at next-to-next-to-leading order (NNLO) accuracy in QCD. We treat the bottom quarks as massless with a non-zero Higgs Yukawa coupling $y_b$. We consider contributions in which the Higgs boson couples directly to bottom quarks, i.e.~our predictions are accurate to order $\mathcal{O}(\alpha_s^3 y_b^2)$. We calculate the various components needed to construct the NNLO contribution, including an independent calculation of the two-loop amplitudes. We compare our results for the two-loop amplitudes to an existing calculation finding agreement. We present additional checks on our two-loop expression using the known infrared factorization properties as the emitted gluon becomes soft or collinear. We use our results to construct a Monte Carlo implementation of \hbbj and present jet rates and differential distributions in the Higgs rest frame using the Durham jet algorithm.}
\maketitle

\flushbottom

\section{Introduction} 

The discovery of the Higgs boson~\cite{Aad:2012tfa,Chatrchyan:2012xdj} has set a large part of the agenda in high energy physics for the foreseeable future. 
Of primary concern is the need to determine the properties of the Higgs boson in relation to the predictions of the Standard Model (SM). 
This is mainly achieved through measurements of the couplings of the Higgs boson to the other SM particles and the Higgs coupling to 
itself. The Higgs self-coupling is of particular interest, since it is intimately linked to the electroweak symmetry breaking potential, the form of which 
is still unconstrained through measurements of the Higgs mass alone (although its remaining properties are predicted in the SM). Any 
additional physics beyond the Standard Model (BSM) could lead to significant changes in the shape of the electroweak symmetry breaking potential, and thus
lead to deviations from the SM predictions. 

Measuring the properties of the Higgs boson is an ongoing task. In regards to that, the LHC has already achieved a remarkable precision with existing Run II measurements and will significantly improve upon these results over the course of the next decade. Plans are afoot for future colliders beyond the LHC (FCs) and a particularly appealing prospect regarding Higgs precision physics is the construction of a lepton collider. Due to the clean experimental conditions, future lepton colliders should be able to probe the properties of the Higgs boson down to per-mille level accuracy~\cite{Baer:2013cma,Gomez-Ceballos:2013zzn,Abada:2019zxq}.

The Higgs boson decays predominately to bottom quark pairs $(b\overline{b}$), and therefore a large part of the experimental program at the LHC and putative FCs 
consists in measuring the properties of this decay. At the LHC the \hbb process can be accessed through associated production channels $pp\rightarrow VH $ followed by a subsequent
\hbb decay~\cite{Aaboud:2018zhk,Sirunyan:2018kst} or directly, by using jet substructure techniques and by looking in the high-$p_T$ $H+j$ channel~\cite{Sirunyan:2017dgcx}, where the backgrounds can be controlled to such a level 
as to make this measurement a possibility. In both situations precise predictions are mandatory to ensure that theoretical calculations have a similar or smaller uncertainty than 
the experimental counterparts. This will become even more pressing at an FC, for which historical measurements from LEP for $Z/\gamma^* \rightarrow$ jets 
already show that the level of experimental uncertainty will be very small indeed. 

Given its importance for LHC physics, the study of Higgs plus multi-parton production has received significant theoretical attention over the last couple of decades. 
Working within the effective field theory, in which the top quark is treated as infinitely heavy, the production of a Higgs through gluon fusion is known to N$^3$LO in QCD~\cite{Anastasiou:2015ema,Anastasiou:2016cez}. Recently, differential predictions at this order have been computed using the method of $Q_T$ subtraction~\cite{Catani:2007vq,Cieri:2018oms} and analytically for the rapidity distribution \cite{Dulat:2017prg,Dulat:2018bfe}. In order to compute $pp\rightarrow H$ differentially at N$^3$LO, $pp\rightarrow H+j$ 
must be available at NNLO, $pp\rightarrow H+2j$ at NLO, and $pp\rightarrow H + 3j$ at LO. These computations have all been performed~\cite{Boughezal:2013uia,Chen:2014gva,Campbell:2006xx}\footnote{Indeed, $H+3j$ is also available at NLO in QCD~\cite{Cullen:2013saa}.}. Of particular note for this work is the calculation of $pp\rightarrow H+j$ at NNLO, which requires the analytic computation of $H\rightarrow 3$ partons in the EFT~\cite{Gehrmann:2011aa}. The related process in which the Higgs boson decays to three partons via a tree-level coupling to $b$-quarks has been less well-studied in the literature. Attention has naturally been focused on the \hbb process which has been studied at NLO~\cite{Braaten:1980yq} and NNLO~\cite{Anastasiou:2011qx,DelDuca:2015zqa,Bernreuther:2018ynm}, and inclusively is known to $\mathcal{O}(\alpha_s^4)$~\cite{Baikov:2005rw}. No complete NNLO prediction for $H \to b \overline{b} j$ is available, although a calculation of the two-loop amplitudes has been presented~\cite{Ahmed:2014pka}. 

The aim of this paper is twofold. Firstly, we perform an independent computation of the  two-loop amplitudes for \hbbg which have been presented in the literature in Ref.~\cite{Ahmed:2014pka}. Secondly, we use these results to produce a NNLO Monte Carlo code for the \hbbj process. The primary goal is to establish whether we can effectively integrate out the additional jet at NNLO. By successfully doing so, we open up the possibility of studying \hbb decay at N$^3$LO.  We perform this calculation in a companion paper~\cite{Mondini:2019gid}.

Our paper proceeds as follows. In Section \ref{sec:calc} we give a general overview of the calculation, while a detailed discussion of our two-loop computation is presented in Section \ref{sec:twoloop}. We discuss the results of our Monte Carlo implementation of $H \to b \overline{b} j$ in Section \ref{sec:results}. After drawing our conclusions, we present the full analytic results of our two-loop amplitudes in the appendix.

\section{Overview of the calculation} 
\label{sec:calc}

\subsection{General overview} 

\begin{figure}
\begin{center}
\includegraphics[width=12cm]{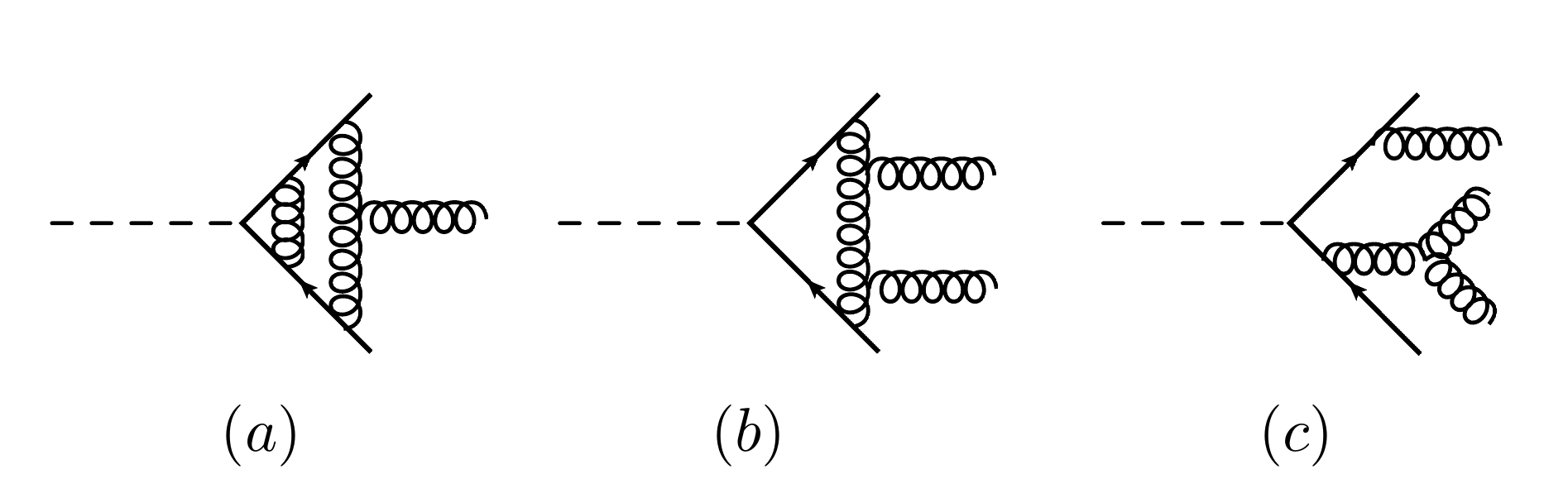}
\caption{Representative Feynman diagrams for the $H \to b \overline{b} j$ process at NNLO.}
\label{fig:hbbgnnlo} 
\end{center}
\end{figure}

In this paper we consider the decay of a Higgs boson to a bottom quark pair and an additional jet at NNLO in QCD. In perturbation theory up to NNLO the partial decay width 
is expanded as follows:
\begin{eqnarray}
\Gamma^{{\rm{NNLO}}}_{H\rightarrow b\overline{b}j} = \Gamma^{{\rm{LO}}}_{H\rightarrow b\overline{b}j} + \Delta \Gamma^{{\rm{NLO}}}_{H\rightarrow b\overline{b}j} + \Delta \Gamma^{{\rm{NNLO}}}_{H\rightarrow b\overline{b}j} \, .
\end{eqnarray}
The above formula introduces the notation we will use in this paper: $\Gamma^{X}_{H\rightarrow b\overline{b}j}$ defines the partial width at order $X$ in perturbation theory, while $\Delta \Gamma^{X}_{H\rightarrow b\overline{b}j}$ defines 
the coefficient which enters the expansion for the first time at this order. 
Representative 
Feynman diagrams for our NNLO calculation are shown in Fig.~\ref{fig:hbbgnnlo}. Specifically, at NNLO we need to compute two-loop amplitudes for $H\rightarrow b\overline{b}g$, one-loop amplitudes 
for \hbbgg and \hbbqq (including identical-quark terms $H\rightarrow b\overline{b}b\overline{b}$), and tree-level amplitudes for $H\rightarrow b\overline{b}ggg$, $H\rightarrow b\overline{b}q\overline{q}g$, and $H\rightarrow b\overline{b}b\overline{b}g$. 

Radiative corrections to the $H\rightarrow b\overline{b}$ decay were first studied nearly forty years ago~\cite{Braaten:1980yq}, when it was shown that 
there are sizable differences between calculations in the ``massless theory'', in which the $b$-quark mass is dropped in the phase space and 
kinematics but kept in the $b$-quark Yukawa coupling, and in the full theory, in which the $b$-quark mass is retained throughout. These differences
were shown to be primarily due to logarithms of the form $\log{(m_b^2/m_H^2)}$. It was also discussed how these effects can be reinstated in the massless theory by running the $b$-quark mass 
in the Yukawa coupling.  Using the $b$-quark mass evolved to the Higgs scale in the massless theory results in much smaller differences between the two theories. This was shown explicitly in Ref.~\cite{Chetyrkin:1996sr}, where the inclusive decay rate $\Gamma_{H\rightarrow b\overline{b}}$ was computed up to order $\mathcal{O}(\alpha_s^3)$ including power-suppressed corrections of the form $m_b^2(m_H)/m_H^2$ up to order $\mathcal{O}(\alpha_s^2)$. The numerical evaluation of the decay rate shows that at each order in $\alpha_s$ the mass corrections are at the per-mille level relative to the massless contribution at the same order and that they are also smaller than the massless corrections at the next order in perturbation theory. It is therefore advantageous and theoretically convenient to work in the massless limit, due to the reduced complexity of higher-order Feynman diagrams. In the massless theory 
the inclusive partial width for the $H\rightarrow b\overline{b}$ decay channel is currently known to an impressive $\mathcal{O}(\alpha_s^4)$ accuracy~\cite{Baikov:2005rw}. The form factor for \hbb at three loops is also 
known~\cite{Gehrmann:2014vha}, so that, once a NNLO calculation of \hbbj is complete, all of the component pieces for \hbb at N$^3$LO are available. 

\begin{figure}
\begin{center}
\includegraphics[width=8cm]{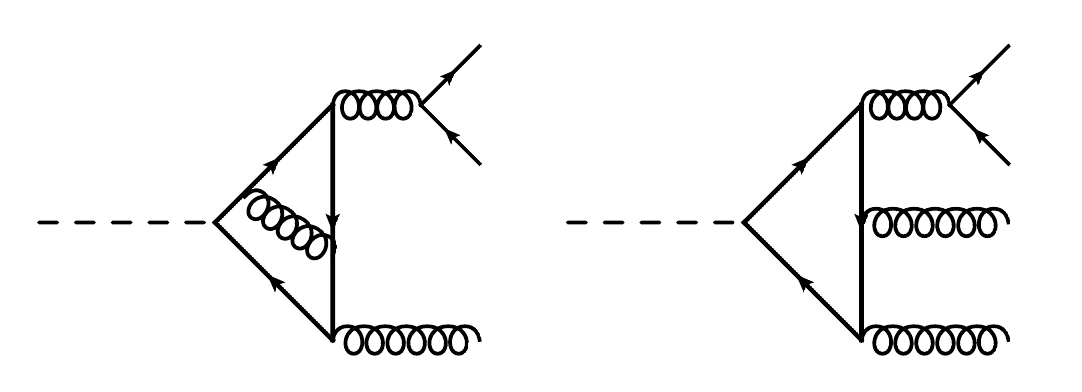}
\caption{Examples of Feynman diagrams that do not enter our calculation at NNLO. }
\label{fig:hbbgnnlo_not} 
\end{center}
\end{figure}

In this paper we will therefore work in the massless theory in which the $b$-quark mass is dropped from the phase space and kinematics, but kept in the Yukawa coupling with 
the $b$-quark mass run to the Higgs scale. As mentioned above, a result of the massless theory assumption is that it simplifies the calculation by reducing the number of Feynman diagrams which must be included at one and two loops. 
We refer in particular to diagrams in which the Higgs boson couples indirectly to the $b$-quarks, for which example topologies are shown in Fig.~\ref{fig:hbbgnnlo_not}.
At $\mathcal{O}(\alpha_s^3)$ these diagrams interfere with the respective tree-level amplitudes for \hbbg and \hbbgg for the two-loop and one-loop calculations respectively. A simple helicity argument 
indicates that these interference terms are zero. In the \hbbg and \hbbgg tree-level amplitudes the scalar Higgs boson couples directly to the two (massless) quarks, which therefore must have identical helicity assignments (both positive or negative). On the other hand, the diagrams in which the Higgs couples implicitly to the $b$ quarks as shown in  Fig.~\ref{fig:hbbgnnlo_not} always result in the final-state $b\overline{b}$ pair coupling directly to a gluon. This vertex requires that the fermions have opposite helicities, and therefore there is no combination that allows non-zero interference terms to exist, resulting in no net contribution from these diagrams at NNLO (the \hbbgg box squared would first enter at $\mathcal{O}(\alpha_s^4)$). 

A slight subtlety arises when we consider the one-loop triangle diagram in which the Higgs boson couples indirectly to the bottom quarks (i.e the left diagram in Fig.~\ref{fig:hbbgnnlo_not} with no additional gluon exchanged in the loop). This diagram would self-interfere at $\mathcal{O}(\alpha_s^3)$ and is therefore not excluded from our NNLO calculation by the argument presented above. However, the trace over the fermion loop 
for this diagram contains five $\gamma$ matrices and hence this term vanishes in the massless theory. In order for this diagram to give a non-zero contribution, the quark mass must be retained in the loop. 
This is the case when the loop particle is a top quark, and hence there exists a top Yukawa contribution which first enters at $\mathcal{O}(\alpha_s^3)$ in our calculation. Schematically, the perturbative 
expansion of the decay width $\Gamma^{{\rm{NNLO}}}_{H \to b\overline{b} j}$ in the full theory is of the form:
\begin{align}
\Gamma^{{\rm{NNLO}}}_{H \to b\overline{b} j} &\sim \as y_b^2 \,A_b + \as^2 \left( y_b^2 \,B_b + y_t y_b \,B_{tb} \right) \notag \\ &\quad + \as^3 \left( y_b^2 \,C_b + y_t^2 \,C_t + y_t y_b \,C_{tb} \right) +\mathcal{O}(\alpha_s^4) \, ,
\end{align}
where $y_b$ and $y_t$ are the bottom and top Yukawa couplings respectively. From the arguments given above it is clear that in the full theory the interference terms $y_t y_b \,B_{tb}$ and $ y_t y_b \,C_{tb}$ are suppressed by the bottom-quark mass (since a helicity flip is needed to make a non-zero interference term). However, since the top Yukawa coupling is large, these mixed terms are of phenomenological relevance. Specifically, in an effective theory in which the top-quark loop is integrated out, the term $y_t y_b \,B_{tb}$ contributes to around 30\% of the $\mathcal{O}(\as^2)$ coefficient \cite{Primo:2018zby}. For our theoretical setup, the mixed term $B_{tb}$ and $C_{tb}$ are exactly zero. In addition, at $\mathcal{O}(\as^3)$ the pure top contribution $ y_t^2 \,C_t$ mentioned above needs to be included. Indeed, while formally this term enters the perturbative expansion as a one-loop squared contribution, the higher-order corrections are known to be large (and well-studied in the EFT approach). This means that for a good phenomenological description higher-order terms proportional to $y_t^2$ should be included as well. The IR properties of this piece are further complicated by the presence of collinear singularities as the $b\overline{b}$ pair becomes unresolved (in the massless theory) since this piece factors onto a different LO term ($H\rightarrow gg$). In this paper we drop the $y_t^2$ term for two reasons. Firstly, we are interested in the theoretical computation of the $y_b^2$ terms (which is new), while the study of the $y_t^2$ contribution has received significant attention in the literature through the various studies of $H+j$ at the LHC. Secondly, we wish to use this computation to perform the N$^3$LO calculation of the $y_b^2$ terms for $H \rightarrow b\overline{b}$. We leave the inclusion of the top Yukawa contributions to a future study, while we remind the reader that these contributions should be included before a complete phenomenological study is performed.

\subsection{$N$-jettiness slicing} 

In order to regulate the IR divergences present in our NNLO calculation we employ the $N$-jettiness slicing method~\cite{Gaunt:2015pea,Boughezal:2015dva}. Since there are three partons in the final 
state at LO we use the 3-jettiness variable $\tau_3$ to separate our calculation into two pieces. For a parton-level  event the 3-jettiness variable \cite{Stewart:2010tn} is defined as follows: 
\begin{eqnarray}
\tau_3 = \sum_{j=1,m} \min_{i=1,2,3} \left\{ \frac{2 q_i \cdot p_{j}}{Q_i}\right\} \, ,
\end{eqnarray}
where the index $j$ runs over the $m$ partons in the phase space (with momenta $p_j$), while $q_i$ represent the momenta of the three most energetic jets, clustered in our case with the Durham jet algorithm~\cite{Brown:1990nm,Catani:1991hj}. $Q_i$ are the hard scales in the process, which are typically taken to be $Q_i = 2E_i$ with $E_i$ the energy of the $i$-th jet. We then introduce a variable \tauc that separates the phase space into two regions. The region $\tau_3 < $\tauc contains all of the doubly-unresolved regions of phase space and here the partial width can be approximated with the following convolution, derived from SCET \cite{Stewart:2010tn,Stewart:2009yx}: 
\begin{eqnarray}
\Gamma_{H\rightarrow 3 j} \left( \tau_3 < \taucm \right) \approx \int \prod_{i=1}^{3} \mathcal{J}_i \otimes \mathcal{S} \otimes \mathcal{H} + \mathcal{O}(\taucm) \, .
\label{eq:bewtau}
\end{eqnarray}
In the above equation the terms $\mathcal{J}_i$ correspond to the jet functions which describe collinear emissions, $\mathcal{S}$ denotes the soft function for three colored partons, and $\mathcal{H}$ is the process-specific hard function. The explicit expressions for the jet functions $\mathcal{J}_i$ needed for our NNLO computation can be found in Ref.~\cite{Becher:2006qw}.
For the soft function, we use the results for the 1-jettiness soft function with arbitrary kinematics computed in Ref.~\cite{Campbell:2017hsw} (see also Ref.~\cite{Boughezal:2015eha}). The calculation of the hard function for this process is one of the primary 
aims of this paper and is discussed in Section~\ref{sec:twoloop}. In order for the approximate form of the partial width in Eq.~\eqref{eq:bewtau} to be accurate, \tauc should be taken 
as small as possible to minimize the power corrections which vanish in the limit \tauc$\rightarrow 0$. 

\subsection{The $\tau_3 > \taucm$ contribution} 

Since any doubly-unresolved contribution resides in the region $\tau_3 < \taucm$, the region $\tau_3 > \taucm$ corresponds to the NLO calculation of $H\rightarrow b\overline{b}jj$. The methods to compute one-loop expressions are by now well-established so we do not spend significant time on them here. In this section we limit ourselves to a brief description of the computation. 
One-loop amplitudes are computed analytically using the generalized unitarity approach~\cite{Bern:1994zx}. Specifically, quadruple cuts are used to compute box coefficients~\cite{Britto:2004nc}, triple cuts are used to compute the triangle coefficients~\cite{Forde:2007mi}, double cuts are used to compute bubble coefficients~\cite{Mastrolia:2009dr}, and the rational pieces are computed using $d$-dimensional unitarity techniques as outlined in Ref.~\cite{Badger:2008cm}. Our calculation is checked numerically using the $d$-dimensional unitarity algorithm presented in Ref.~\cite{Ellis:2008ir}.  The resulting expressions are rather compact, with a similar level of complexity to the $H \to gggg$ amplitudes presented in Ref.~\cite{Badger:2009hw}. 
Tree-level amplitudes are computed using the BCFW recursion relations~\cite{Britto:2005fq} and all tree-level amplitudes present in the calculation have been checked against Madgraph~\cite{Alwall:2011uj}. Finally, IR divergences in the NLO calculation are regulated using Catani-Seymour dipole subtraction~\cite{Catani:1996vz}.

\section{Hard function for $H\rightarrow b\overline{b}g$ at NNLO}
\label{sec:twoloop}

In this section we describe the calculation of the hard function $\mathcal{H}$ of Eq.~\eqref{eq:bewtau} for the process $H\rightarrow b\overline{b}g$ at NNLO accuracy. We define the hard function as a perturbative series in powers of the renormalized strong coupling $\as \equiv \as(\mu)$ at the renormalization scale $\mu$:
\begin{align}
\mathcal{H} = \mathcal{H}_{\text{LO}} + \left (\frac{\alpha_s}{2\pi} \right ) \mathcal{H}_{\text{NLO}} + \left (\frac{\alpha_s}{2\pi} \right )^2 \mathcal{H}_{\text{NNLO}} + \mathcal{O}(\alpha_s^3) \, .
\end{align}
The LO, NLO, and NNLO coefficients of the hard function are
\begin{align}
\mathcal{H}_{\text{LO}} &= \mathcal{M}^{(0),\text{ren}} {\mathcal{M}^{(0),\text{ren}}}^{*} \label{lohard} \\
\mathcal{H}_{\text{NLO}} &= 2 \,\text{Re}\left (\mathcal{M}^{(1),\text{ren}} {\mathcal{M}^{(0),\text{ren}}}^{*} \right ) \\
\mathcal{H}_{\text{NNLO}} &= \mathcal{M}^{(1),\text{ren}} {\mathcal{M}^{(1),\text{ren}}}^{*} + 2 \,\text{Re}\left (\mathcal{M}^{(2),\text{ren}} {\mathcal{M}^{(0),\text{ren}}}^{*} \right ) \label{nnlohard}
\end{align}
where $\mathcal{M}^{(\ell),\text{ren}}$ is the $\overline{\text{MS}}$-renormalized $\ell$-loop amplitude in the notation of Ref.~\cite{Becher:2013vva}. The calculation of $\mathcal{M}^{(\ell),\text{ren}}$ with $\ell=0,1,2$ is described in the following sections.

\subsection{Notation and kinematics}

We consider the decay 
\begin{align*}
H \to b(p_1)\,\bar{b}(p_2)\,g(p_3) \, .
\end{align*}
The Mandelstam invariants for this process are defined as
\begin{align*}
s = (p_1+p_2)^2 > 0 \hspace{1.5cm} t = (p_1+p_3)^2 > 0 \hspace{1.5cm} u = (p_2+p_3)^2 > 0
\end{align*}
and satisfy $s+t+u=m_H^2$ with $m_H$ the mass of the Higgs boson. We also introduce the dimensionless quantities
\begin{align}
x = \frac{s}{m_H^2} \hspace{1cm} y = \frac{t}{m_H^2} \hspace{1cm} z = \frac{u}{m_H^2}
\end{align}
which satisfy $0<x<1$, $0<y<1$, $0<z<1$, and $x+y+z=1$.

We follow the notation introduced in Ref.~\cite{Ahmed:2014pka}, in which the unrenormalized amplitude for $H\rightarrow b\overline{b}g$ is written in terms of two tensor structures:
\begin{align} \label{mintermsofa1a2}
\mathcal{M} = i \,\left (\frac{\overline{\alpha}_s}{2\pi}\right )^{\frac{1}{2}} \frac{\overline{y}_b}{m_H^2}\, \TAij\, \epsilon_{\mu}(p_3) \left [ A_1 \, T_1^{\mu} + A_2 \, T_2^{\mu} \right ] \, ,
\end{align}
where $\overline{\alpha}_s$ is the bare strong coupling constant, $\overline{y}_b$ is the bare  bottom Yukawa coupling, $\TAij$ is the color matrix with gluon color index $a$ and quark indices $i$ and $j$, and $\epsilon_{\mu}(p_3)$ is the gluon polarization vector. Finally, the tensors $T_1^{\mu}$ and $T_2^{\mu}$ are defined as
\begin{align}
T_1^{\mu} &= \bar{u}(p_1) \slsh p_3 \, \gamma^{\mu} \, v(p_2) \notag \\
T_2^{\mu} &= \left [ p_1^{\mu} - \frac{t}{u} p_2^{\mu} \right ] \bar{u}(p_1) \, v(p_2) \, . \label{tensordefeq}
\end{align}
The coefficients $A_m$ ($m=1,2$) have perturbative expansions in powers of $\overline{\alpha}_s$:
\begin{align} \label{acoeffexp}
A_m = A_m^{(0)} + \left (\frac{\overline{\alpha}_s}{2\pi}\right ) A_m^{(1)} + \left (\frac{\overline{\alpha}_s}{2\pi}\right )^2 A_m^{(2)} + \mathcal{O}(\overline{\alpha}_s^3)
\end{align}
where the coefficients $A_m^{(\ell)}$ with $\ell \geq 1$ contain UV and IR divergences which are regularized in $d=4-2\eps$ dimensions.  At any order in $\overline{\alpha}_s$, the coefficients $A_m^{(\ell)}$ are obtained by applying the projectors
\begin{align}
P_1^{\mu} &= -\frac{1}{2 (d-3) t u} T_1^{\mu \dagger} -\frac{1}{2 (d-3) s t} T_2^{\mu \dagger} \notag \\
P_2^{\mu} &= -\frac{1}{2 (d-3) s t} T_1^{\mu \dagger} -\frac{(d-2) u}{2 (d-3) s^2 t} T_2^{\mu \dagger} 
\end{align}
to the appropriate amplitude, namely
\begin{align} \label{howtogetacoeff}
A_m^{(\ell)} = \sum_{\text{pol}} P_m^{\nu}\, \epsilon^{*}_{\nu}(p_3) \mathcal{M}^{(\ell)}
\end{align}
where $\mathcal{M}^{(\ell)}$ is the $\ell$-loop amplitude written, for instance, as the sum of Feynman diagrams. The sum over the polarization states of the external gluon is performed as
\begin{align}
\sum_{\text{pol}} \epsilon^{\mu}(p_3) \epsilon^{\nu *}(p_3) = -g^{\mu \nu} + \frac{p_3^{\mu} q^{\nu} + q^{\mu} p_3^{\nu}}{q \cdot p_3}
\end{align}
where $q$ is an auxiliary vector. In our calculation we choose $q=p_1$.

\subsection{Calculation}

We now discuss the calculation of the coefficients $A_m$ to second order.  We generate the tree-level, one-loop, and two-loop Feynman diagrams using FeynArts \cite{Hahn:2000kx}. At tree level, by applying Eq.~\eqref{howtogetacoeff} and by carrying out the trace calculations in $d$ dimensions we directly obtain $A_m^{(0)}$. At one loop and two loops, after using Eq.~\eqref{howtogetacoeff} the coefficients $A_m^{(1)}$ and $A_m^{(2)}$ are written in terms of scalar one-loop and two-loop integrals respectively. We reduce them to an irreducible set of master integrals (MIs) using the programs Kira \cite{Maierhoefer:2017hyi} and LiteRed \cite{Lee:2013mka}. The topologies needed to reduce all integrals appearing in the calculation are the same as those presented in Eqs.~(3.2)-(3.5) of Ref.~\cite{Ahmed:2014pka}.

At the one-loop level, there are two master integrals, namely the bubble and the box integral. Their explicit results are presented in Appendix A of Ref.~\cite{Garland:2001tf}, where in particular the result for the box integral is given as a series in the regulator $\epsilon$ and in terms of HPLs \cite{Remiddi:1999ew} and two-dimensional HPLs (2dHPLs) \cite{Gehrmann:2000zt,Gehrmann:2001ck}.

At two loops, all required master integrals are known in the literature and can be divided into three groups: planar integrals, whose results are presented in Ref.~\cite{Gehrmann:2000zt}, non-planar integrals, computed in Ref.~\cite{Gehrmann:2001ck}, and products of two one-loop integrals. As in the case of the one-loop box integral, the results for the two-loop planar and non-planar integrals are expressed as Laurent series in $\epsilon$ and in terms of HPLs and 2dHPLs. Furthermore, following the discussion in Section (3.3) of Ref.~\cite{Garland:2001tf}, we observe that in our calculation each master integral can be present in up to six kinematic configurations (i.e.~with all possible permutations of the independent external momenta $p_1,p_2,p_3$). This means that, after substituting the explicit results of the MIs, our results for the coefficients $A_m^{(\ell)}$ initially contain HPLs with three arguments ($x$, $y$, or $z$) and 2dHPLs with six combinations ($x$, $y$, or $z$ in the index vector and in the argument). In order to simplify our expressions, we can express all HPLs and 2dHPLs appearing in the calculation in terms of HPLs and 2dHPLs belonging to one unique kinematic configuration. Following Refs.~\cite{Garland:2001tf,Gehrmann:2000zt,Gehrmann:2001ck}, we choose 2dHPLs of argument $y$ and index $z$ and HPLs of arguments $y$ and $z$ as the unique set.

One way of obtaining the relations needed to convert all ``spurious'' HPLs to a unique set is by exploiting their integral representation and applying interchange of arguments formulae as described in Refs.~\cite{Garland:2001tf,Gehrmann:2000zt}. In this work we proceed in a slightly different way, following the work on multiple polylogarithms (MPLs), of which HPLs and 2dHPLs are examples, of Ref.~\cite{Duhr:2014woa}. In Ref.~\cite{Duhr:2014woa} it is shown that MPLs form a Hopf algebra and that a coproduct on MPLs can be defined. The coproduct allows one to systematically decompose MPLs of any weight into MPLs of lower weights. Since at weight 1 it is trivial to convert HPLs and 2dHPLs of different arguments and/or indices to a unique set, we can apply the coproduct with a bottom-up approach to find relations between HPLs and 2dHPLs of different kinematic configurations at any weight. In our case we derive all the relations required to reduce HPLs and 2dHPLs of up to weight 4 to the chosen set using the coproduct method. We also use GiNaC to numerically evaluate the 2dHPLs for checking purposes.

\subsection{$\overline{\text{MS}}$-renormalized amplitudes}

We now construct the $\overline{\text{MS}}$-renormalized amplitudes $\mathcal{M}^{(\ell),\text{ren}}$ that are needed for the hard function computation at NNLO accuracy. Through Eq.~\eqref{mintermsofa1a2} this is equivalent to constructing the $\overline{\text{MS}}$-renormalized coefficients $A_m^{(\ell),\text{ren}}$.

\subsubsection{UV renormalization}

We start by removing the UV divergences from the coefficients $A_m^{(\ell)}$ computed in the previous section. We renormalize the bare strong coupling constant and Yukawa coupling by performing the replacements
\begin{align}
\overline{\alpha}_s &\to \as\, S_{\epsilon} \, Z_{\alpha} \label{asren} \\
\overline{y}_b &\to y_b \, Z_y \label{ybren}
\end{align}
with $S_{\epsilon} = \frac{\exp{(\epsilon \gamma_E)}}{(4\pi)^{\epsilon}}$, $\as \equiv \as(\mu)$ and $y_b \equiv y_b(\mu)$ at the renormalization scale $\mu$. The renormalization factors are given by
\begin{align}
Z_{\alpha} &= 1 + \left (\frac{\as}{2\pi}\right ) r_1 + \left (\frac{\as}{2\pi}\right )^2 r_2 + \mathcal{O}(\alpha_s^3) \label{zalphaseries} \\
Z_y &= 1 + \left (\frac{\as}{2\pi}\right ) s_1 + \left (\frac{\as}{2\pi}\right )^2 s_2 + \mathcal{O}(\alpha_s^3) \label{zyseries}
\end{align}
with $r_1$, $r_2$, $s_1$, $s_2$ explicitly defined in Appendix \ref{renoirformulae}. By inserting Eqs.~\eqref{asren} and \eqref{ybren} into Eqs.~\eqref{mintermsofa1a2} and \eqref{acoeffexp}, we obtain the UV-finite coefficients $A_m^{(\ell),\text{UV-fin}}$:
\begin{align}
A_m^{(0),\text{UV-fin}} &= A_m^{(0)} \\
A_m^{(1),\text{UV-fin}} &= S_{\epsilon} A_m^{(1)} + \left (s_1+\frac{r_1}{2}\right ) A_m^{(0)} \\
A_m^{(2),\text{UV-fin}} &= S_{\epsilon}^2 A_m^{(2)} + \left (s_1+\frac{3 r_1}{2}\right ) S_{\epsilon} A_m^{(1)} \notag \\ &\quad + \left (s_2 + \frac{r_1 s_1}{2} +\frac{r_2}{2} -\frac{r_1^2}{8} \right ) A_m^{(0)} \, .
\end{align}

\subsubsection{IR subtraction and conversion to $\overline{\text{MS}}$ scheme}

In order to obtain the hard function we remove the explicit soft and collinear divergences from the UV-renormalized coefficients. The IR structure of one-loop and two-loop QCD amplitudes is universally known \cite{Catani:1998bh} and can be written using Catani's subtraction operators $I^{(\ell)}(\epsilon)$. The finite coefficients $A_m^{(\ell),\text{fin}}$ are defined as
\begin{align}
A_m^{(0),\text{fin}} &= A_m^{(0),\text{UV-fin}} \\
A_m^{(1),\text{fin}} &= A_m^{(1),\text{UV-fin}} - I^{(1)}(\epsilon) A_m^{(0),\text{UV-fin}} \\
A_m^{(2),\text{fin}} &= A_m^{(2),\text{UV-fin}} - I^{(1)}(\epsilon) A_m^{(1),\text{UV-fin}} - I^{(2)}(\epsilon) A_m^{(0),\text{UV-fin}} \, .
\end{align}
The explicit expressions of the subtraction operators for \hbbg can be found in Appendix \ref{renoirformulae}. In Appendix \ref{formulaehbbg} we show the complete results for the coefficients $A_m^{(\ell),\text{fin}}$. Specifically, following the notation of Eq.~(4.4) of Ref.~\cite{Ahmed:2014pka}, we write the coefficients as
\begin{align} \label{bcoeffdef}
A_m^{(\ell),\text{fin}} &= \sum_{n=0}^\ell A_m^{(0),\text{fin}} \,\mathcal{B}_{m;n}^{(\ell)} \,\text{ln}^n \! \left (-\frac{m_H^2}{\mu^2}\right )
\end{align}
with the coefficients $A_m^{(0),\text{fin}}$ and $\mathcal{B}_{m;n}^{(\ell)}$ presented in Appendix \ref{formulaehbbg}.

Finally, following the discussion in Section (2.1) of Ref.~\cite{Becher:2013vva}, we obtain the $\overline{\text{MS}}$-renormalized coefficients $A_m^{(\ell),\text{ren}}$ in the following way:
\begin{align}
A_m^{(0),\text{ren}} &= A_m^{(0),\text{fin}} \\
A_m^{(1),\text{ren}} &= A_m^{(1),\text{fin}} + \mathcal{C}_0 A_m^{(0),\text{fin}} \label{mscoeff1} \\
A_m^{(2),\text{ren}} &= A_m^{(2),\text{fin}} + \mathcal{C}_0 A_m^{(1),\text{fin}} + \mathcal{C}_2 A_m^{(0),\text{fin}} \label{mscoeff2}
\end{align}
where $\mathcal{C}_0$ and $\mathcal{C}_2$ are defined in Appendix \ref{renoirformulae}. By using Eqs.~\eqref{lohard}-\eqref{nnlohard} and \eqref{mintermsofa1a2} we obtain the hard function at NNLO accuracy. Explicitly, the interferences are constructed as follows:
\begin{align}
\mathcal{M}^{(m),\text{ren}} {\mathcal{M}^{(n),\text{ren}}}^{*} &= \mathcal{N}_{LO} \Big ( 4 y z \, A_1^{(m),\text{ren}} {A_1^{(n),\text{ren}}}^{*} + \frac{2 x^2 y}{z} A_2^{(m),\text{ren}} {A_2^{(n),\text{ren}}}^{*} \notag \\ &\quad -2 x y \, A_1^{(m),\text{ren}} {A_2^{(n),\text{ren}}}^{*} -2 x y\, A_2^{(m),\text{ren}} {A_1^{(n),\text{ren}}}^{*} \Big )
\end{align}
where $\mathcal{N}_{LO} = \left (\frac{\alpha_s}{2\pi} \right ) y_b^2 \, N_c C_F$.

\subsection{Comparison with existing results}

We can compare our results for the coefficients $A_m^{(\ell),\text{fin}}$ up to $\ell=2$ with the existing results in the literature \cite{Ahmed:2014pka}. At tree level the agreement is trivial. Since we defined the tensors $T_1^{\mu}$ and $T_2^{\mu}$ as in Eq.~\eqref{tensordefeq}, the relation between our coefficients $A_m^{(0),\text{fin}}$ and the corresponding ones of Ref.~\cite{Ahmed:2014pka} (here called $A_{1;\,\text{lit}}^{(0)}$ and $A_{2;\,\text{lit}}^{(0)}$) is
\begin{align}
A_{1;\,\text{lit}}^{(0)} = \frac{2 i}{z} \, A_2^{(0),\text{fin}} \hspace{0.5cm} \text{and} \hspace{0.5cm} A_{2;\,\text{lit}}^{(0)} =  i \, A_1^{(0),\text{fin}} \, .
\end{align}
At one and two loops we can compare the coefficients $\mathcal{B}^{(\ell)}_{m;n}$ with those presented in Appendix B and C of Ref.~\cite{Ahmed:2014pka}. Since the coefficients $\mathcal{B}^{(\ell)}_{m;n}$ have been rescaled by the tree-level coefficients, we only need to swap $\mathcal{B}^{(\ell)}_{1;n} \leftrightarrow \mathcal{B}^{(\ell)}_{2;n}$ to match our notation. We find complete agreement for all coefficients at one-loop level\footnote{After adjusting for a small typo (i.e.~changing $12 H(0,2;y) \to 2 H(0,2;y)$ in the last line of $\mathcal{B}^{(1)}_{1;0}$) and dividing the literature results by an overall factor of 2 (since in the literature results the expansion parameter is $\alpha_s/4\pi$).} and at two-loop level\footnote{After taking into account the different definition of the Mandelstam invariants $t$ and $u$ of Ref.~\cite{Ahmed:2014pka} and after adjusting the literature results by an overall factor of 4.}. The agreement at two loops is explicitly shown in table \ref{tab1} where we perform a numerical comparison between the two sets of results for a random phase-space point.

\begin{table}
\centering
\begin{tabular}{|c|c|c|}
\hline
\rule{0pt}{2ex}{Coefficient} & {Ref.~\cite{Ahmed:2014pka}} & {Our result}\\
\hline
$\mathcal{B}^{(2)}_{1;2}$ & $15.1770833333333$ & $15.1770833333333$\\
\hline
$\mathcal{B}^{(2)}_{1;1}$ & $-61.1367801007938$ & $-61.1367801007938$\\
\hline
$\mathcal{B}^{(2)}_{1;0}$ & $77.6770380202061$ & $77.6770380202060$\\
\hline
$\mathcal{B}^{(2)}_{2;2}$ & $15.1770833333333$ & $15.1770833333333$\\
\hline
$\mathcal{B}^{(2)}_{2;1}$ & $-54.6784467674605$ & $-54.6784467674605$\\
\hline
$\mathcal{B}^{(2)}_{2;0}$ & $74.2152337563907$ & $74.2152337563904$\\
\hline
\end{tabular}
\caption{Numerical comparison between our two-loop results and those of Ref.~\cite{Ahmed:2014pka} for $y=0.19$ and $z=0.67$ after adjusting for an overall $1/4$ factor.}
\label{tab1}
\end{table}

\subsection{Factorization properties of the two-loop amplitude}

Although we established agreement between our two-loop amplitude and an existing result in the literature, both share certain similarities (namely an expansion in the same master integrals). 
We therefore initiate further testing of our calculation by investigating the analytic structure of our result in the limits in which one of the partons becomes unresolved. Such a check was not detailed previously. 
We do so by checking that our two-loop amplitude correctly reproduces the known IR factorization properties of QCD~\cite{Li:2013lsa,Badger:2004uk,Duhr:2014nda} when the external gluon becomes either soft or collinear to one of the quarks. We note that a further by-product of this check is a confirmation of the computed factorization limits of QCD for the soft~\cite{Li:2013lsa} and collinear~\cite{Badger:2004uk} limit.

\subsubsection{Soft-gluon limit}

In the limit of soft gluon, the momentum of the gluon vanishes, i.e.~$p_3\to 0$ which implies that $y,z \to 0$ simultaneously. The soft-gluon limit at two loops reads:
\begin{align}
2 \,\text{Re}\left (\mathcal{M}^{(2)}_{H \to b \bar{b} g} \mathcal{M}^{(0)*}_{H \to b \bar{b} g}\right ) &\to S^{(2)}_{H \to b \bar{b} g} = 2 \,\text{Re} \Big (
S^{(0)}(y,z) \mathcal{M}^{(2)}_{H \to b \bar{b}} \mathcal{M}^{(0)*}_{H \to b \bar{b}} \notag \\ 
&\quad + S^{(1)}(y,z) \mathcal{M}^{(1)}_{H \to b \bar{b}} \mathcal{M}^{(0)*}_{H \to b \bar{b}} \notag \\
&\quad + S^{(2)}(y,z) \mathcal{M}^{(0)}_{H \to b \bar{b}} \mathcal{M}^{(0)*}_{H \to b \bar{b}} \Big ) \, , \label{softlimiteq}
\end{align}
where the relevant $H \to b \bar{b}$ matrix elements and the soft currents $S^{(0)}(y,z)$, $S^{(1)}(y,z)$, $S^{(2)}(y,z)$ are presented in Appendix \ref{matrixelements}. Using our results for the unrenormalized IR-divergent coefficients $A_m^{(2)}$ we construct the interference $2 \,\text{Re}\left (\mathcal{M}^{(2)}_{H \to b \bar{b} g} \mathcal{M}^{(0)*}_{H \to b \bar{b} g}\right )$ as a series in $\epsilon$ in order to compare it with the known soft limit $S^{(2)}_{H \to b \bar{b} g}$ defined above. Since the soft limit diverges as $(y z)^{-1}$, we multiply both expressions by a factor of $y\,z$. We show the obtained numerical results in table \ref{tab3}. The agreement between the known soft limit and our results is excellent.

\begin{table}
\centering
\begin{tabular}{|c|c|c|c|}
\hline
\rule{0pt}{2ex}{Coefficient} & $ y\,z\,S^{(2)}_{H \to b \bar{b} g}$ & Our result\\
\hline
$\epsilon^{-4}$ & $81.7702729678$ & $81.7702729678$\\
\hline
$\epsilon^{-3}$ & $3818.49680411$ & $3818.49680413$\\
\hline
$\epsilon^{-2}$ & $130763.8079162$ & $130763.8079168$\\
\hline
$\epsilon^{-1}$ & $3.26338843478 \cdot 10^6$ & $3.26338843480 \cdot 10^6$\\
\hline
$\epsilon^{0}$ & $6.52342650778 \cdot 10^7$ & $6.52342650793 \cdot 10^7$\\
\hline
\end{tabular}
\caption{Numerical comparison of our two-loop results with the known soft limit for $y=z=10^{-10}$ and $\mu^2 = \frac{m_H^2}{2}$. An overall factor of $\as^3\,y_b^2$ has been extracted from both results.}
\label{tab3}
\end{table}

\subsubsection{Collinear limit}

In the limit of the gluon becoming collinear to the outgoing quark, the invariant $t$ vanishes which means $y \to 0$ while $z\neq 0$. The collinear limit at two loops reads:
\begin{align}
2 \,\text{Re}\left (\mathcal{M}^{(2)}_{H \to b \bar{b} g} \mathcal{M}^{(0)*}_{H \to b \bar{b} g}\right ) &\to C^{(2)}_{H \to b \bar{b} g} = 
2 \,\text{Re} \Big (
C^{(0)}(y,z) \mathcal{M}^{(2)}_{H \to b \bar{b}} \mathcal{M}^{(0)*}_{H \to b \bar{b}} \notag \\ 
&\quad + C^{(1)}(y,z) \mathcal{M}^{(1)}_{H \to b \bar{b}} \mathcal{M}^{(0)*}_{H \to b \bar{b}} \notag \\
&\quad + C^{(2)}(y,z) \mathcal{M}^{(0)}_{H \to b \bar{b}} \mathcal{M}^{(0)*}_{H \to b \bar{b}} \Big ) \, . \label{collinearlimiteq}
\end{align}
The splitting functions $C^{(0)}(y,z)$, $C^{(1)}(y,z)$, $C^{(2)}(y,z)$ are given in Appendix \ref{matrixelements}. We compare our result for $2 \,\text{Re}\left (\mathcal{M}^{(2)}_{H \to b \bar{b} g} \mathcal{M}^{(0)*}_{H \to b \bar{b} g}\right )$ as a series in $\epsilon$ with $C^{(2)}_{H \to b \bar{b} g}$. We multiply both expressions by a factor of $y$ to remove the leading divergence. The numerical results are shown in table \ref{tab4}. We observe excellent agreement between our result and the known collinear limit.

\begin{table}
\centering
\begin{tabular}{|c|c|c|c|}
\hline
\rule{0pt}{2ex}{Coefficient} & $ y\,C^{(2)}_{H \to b \bar{b} g}$ & Our result\\
\hline
$\epsilon^{-4}$ & $283.156234427$ & $283.156234427$\\
\hline
$\epsilon^{-3}$ & $8122.55721506$ & $8122.55721505$\\
\hline
$\epsilon^{-2}$ & $170379.942318$ & $170379.942317$\\
\hline
$\epsilon^{-1}$ & $2.584146 \cdot 10^6$ & $2.584189 \cdot 10^6$\\
\hline
$\epsilon^{0}$ & $3.09852 \cdot 10^7$ & $3.09870 \cdot 10^7$\\
\hline
\end{tabular}
\caption{Numerical comparison between our two-loop results and the known collinear limit for $y=10^{-12}$, $z=0.23$ and $\mu^2 = \frac{m_H^2}{2}$. An overall factor of $\as^3\,y_b^2$ has been extracted from both results.}
\label{tab4}
\end{table}

\subsection{Summary}

In this section we have presented the computation of the hard function required to construct the $\tau_3 < \taucm$ part of our NNLO calculation. We have compared our calculation to a similar existing result in the literature and found agreement. We have also verified that our expressions reproduce the known soft and collinear limits at this order and are therefore confident in using our results for the phenomenology presented in the subsequent sections of this paper.

\section{Results}
\label{sec:results} 

We have implemented the results discussed in the previous sections into a fully-flexible parton-level Monte Carlo code. Our code is based upon the existing structure of MCFM~\cite{Campbell:1999ah,Campbell:2011bn,Campbell:2015qma,Boughezal:2016wmq}
and could be easily included in a future release of the code. 
Here we present phenomenological results for \hbbj. As outlined in Section \ref{sec:calc}, the $b$-quark mass is set to zero kinematically, but kept in the Yukawa coupling. 
In order to account for some of the effects of the missing $b$-mass terms we evolve the $b$-quark mass to the Higgs scale ($m_H =125$ GeV) using the two-loop running for NLO predictions, and three-loop running for NNLO predictions.
This results in an effective $b$-quark mass of 2.94 GeV at NNLO (for our central scale choice $\mu=m_H$). We also use $G_F = 0.116639 \times 10^{-4}$ GeV$^{-2}$ and $m_W = 80.385$ GeV. We take $\alpha_s(m_Z) = 0.118$ and we run  the coupling at one, two, and three loops for LO, NLO, and NNLO calculations respectively. All results in this paper compute the width in units of MeV. In order to compute rates and distributions for $H \rightarrow b\overline{b}j$, a jet algorithm must be applied. In this paper we will present results using the Durham jet algorithm \cite{Brown:1990nm,Catani:1991hj}, which takes the variable $y_{\rm{cut}}$ as an input variable. Starting at the parton level, the algorithm computes the following quantity for every possible pair of partons $(i,j)$:
\begin{eqnarray}
y_{ij} = \frac{2\, {\rm{min}}(E_i^2,E_j^2)(1-\cos\theta_{ij})}{Q^2}
\end{eqnarray}
where $E_i$ is the energy of parton $i$, $\theta_{ij}$ is the angle between partons $i$ and $j$, and in our case $Q=m_H$. If $y_{ij} < y_{\rm{cut}}$ the pairs are combined into a new object with momentum $p_i+p_j$. The algorithm then repeats until no further clusterings are possible and the remaining objects are classified as jets. These algorithms have been widely used at LEP to study $e^+e^-\rightarrow $ jets, which is the process most similar to our \hbbj calculation. 
Our results are presented in the Higgs rest frame.

\begin{figure}
\begin{center}
\includegraphics[width=11cm]{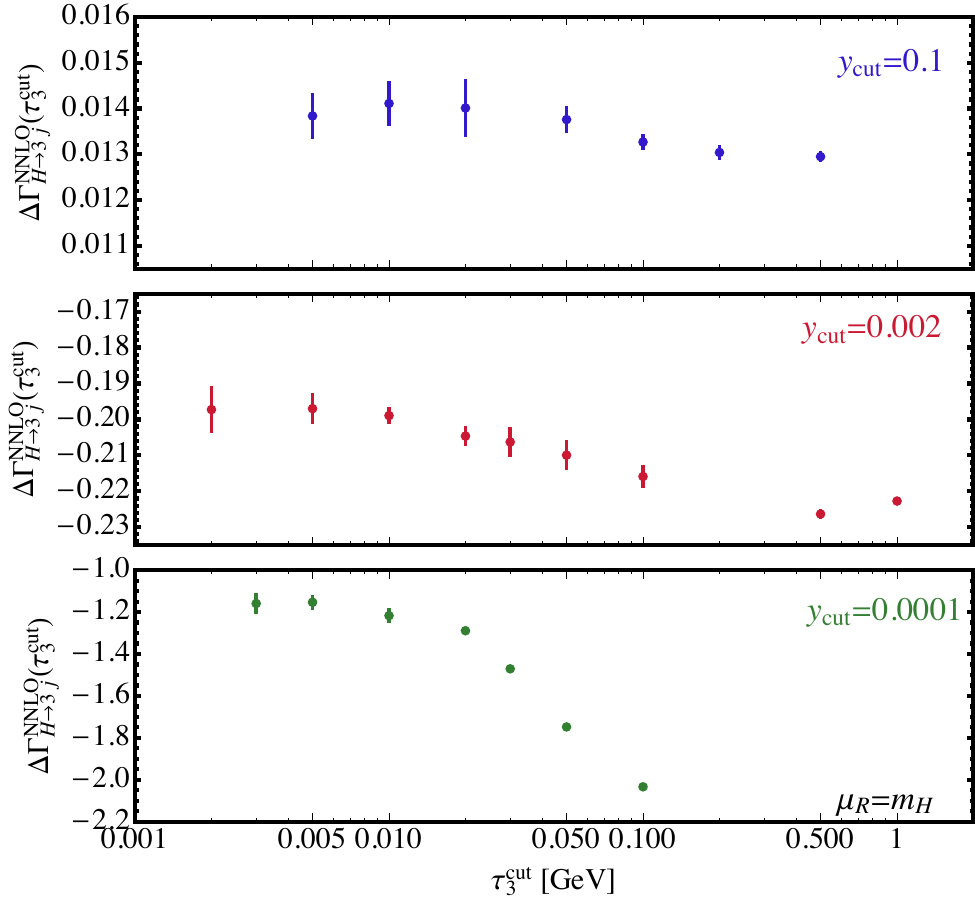}
\caption{The $\tau_{3}^{\rm{cut}}$ dependence of the NNLO coefficient for three different jet definitions.}
\label{fig:taudep} 
\end{center}
\end{figure}

We first validate our calculation by studying the dependence of the NNLO coefficient on the unphysical slicing parameter $\tau_{3}^{\rm{cut}}$. To do so we focus on three representative clustering options corresponding to $y_{\rm{cut}} = 0.1$, 0.002 and $10^{-4}$. These choices span the various regions of interest theoretically and experimentally. The value $y_{\rm{cut}} = 0.1$ is within the perturbative regime, in which the higher-order corrections are expected to be small and agreement with future data should be good (assuming similarity to the NNLO calculations of $e^+e^-\rightarrow$ jets~\cite{GehrmannDeRidder:2008ug,Weinzierl:2008iv}). The second choice $y_{\rm{cut}} = 0.002$ corresponds to the region in which the three-jet rate peaks. Finally, the choice $y_{\rm{cut}} =10^{-4}$ is around the region in which the NNLO three-jet rate turns negative and becomes unphysical (the need for resummation of large $y_{\rm{cut}}$ logarithms has set in long before this value is reached). The final choice is of particular relevance to this paper, since it corresponds to integrating the NNLO calculation with a very weak jet cut.
Creating stable (and slicing-independent) results in this region allows us to test the code in phase-space configurations which correspond to two hard jets and one soft/collinear jet. Such configurations occur copiously in the calculation of \hbb at N$^3$LO (where the soft jet is not required), and therefore establishing our code here is a prerequisite for this computation. 

Our results for the three $y_{\rm{cut}}$ values are presented in Fig.~\ref{fig:taudep}. 
Asymptotic behavior is established in each region, with the dependence on missing power corrections having, as expected, a notable dependence on $y_{\rm{cut}}$. For the larger choices the dependence on \tauc
is rather mild, as the result for the largest value of \tauc is less than 10\% different to that obtained in the asymptotic region (around \tauc $ \le 0.05 $ GeV for $y_{\rm{cut}} = 0.1$ and  \tauc $ \le 0.01 $ for $y_{\rm{cut}} = 0.002$). The dependence on \tauc for \yc $=10^{-4}$ is greater and asymptotic behavior is found for  \tauc $ \le 0.005 $ GeV. We therefore conclude that the power corrections are under control and that our code can be used to make phenomenological predictions. We note in passing that an LHC jet would be clustered using a $k_T$-style algorithm and a jet with around $p_T > 30$ GeV would loosely scale like $\sqrt{m_H^2 \ycm} \sim 30$ GeV, so that the LHC case would look most like our results obtained when $y_{\rm{cut}} \sim 0.1$. In this region we have established that the power corrections are small and under control, and therefore our code could readily be applied to LHC processes such as $pp\rightarrow V(H\rightarrow 3j)$. We leave this study to future work.

\begin{figure}
\begin{center}
\includegraphics[width=7cm]{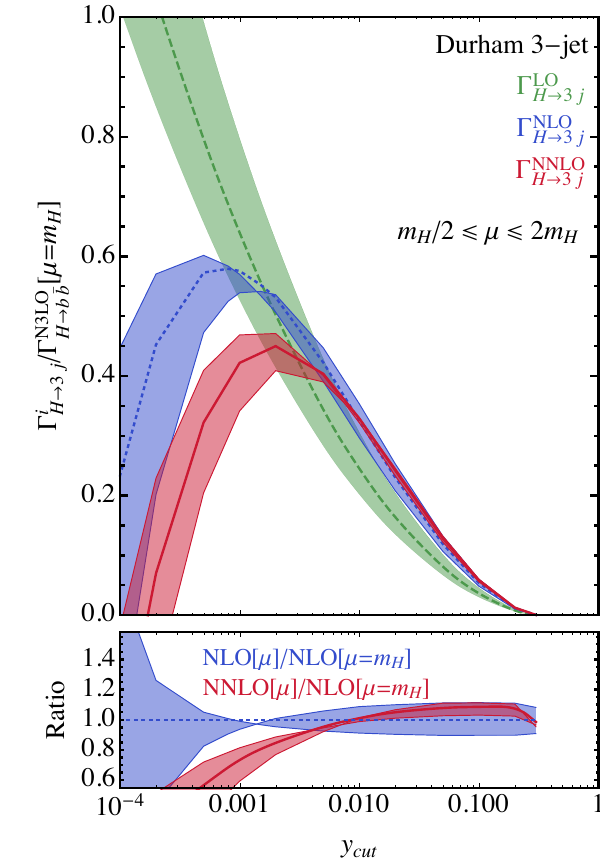}
\caption{The three-jet rate at LO, NLO, and NNLO as a function of $y_{\rm{cut}}$ for the Durham jet algorithm. The renormalization scale is set to $\mu = m_H$.}
\label{fig:jetrateNNLO} 
\end{center}
\end{figure}

In Fig.~\ref{fig:jetrateNNLO} we show the exclusive three-jet rate at LO, NLO, and NNLO as a function of $y_{\rm{cut}}$. We present results for the three-jet rate normalized to the N$^3$LO $H \to b \overline{b}$ inclusive rate \cite{Baikov:2005rw}. In order to make each prediction we have set \tauc $=10^{-2}$ GeV, which is in the asymptotic region for nearly all of the phase space of interest. This choice is slightly too large for the smallest value of \yc studied as discussed in the previous paragraph. However, the error on the coefficient for this choice is around 5\%, which corresponds to a phenomenologically-acceptable $\sim 2\%$ correction on the total fractional jet rate. Our figure can be compared to 
similar results obtained for $e^+e^-\rightarrow$ jets \cite{GehrmannDeRidder:2008ug,Weinzierl:2008iv}. The pattern is broadly the same, with a small positive correction in the large \yc region (around 10\%), which transitions to a decrease in the jet rate for $y_{\rm{cut}} \le 0.01$. The three-jet rate is maximum at around \yc$=0.002$ and then turns over, becoming negative (and hence unphysical) in the region around $10^{-4}$. 
Along with the central scale choice of $\mu=m_H$ we also provide predictions for jet rates obtained with renormalization scales $\mu = \{2,1/2\} \times m_H$. In addition to the implicit dependence in the loop integral expansion, the predictions depend on $\mu$ also through the running of $\alpha_s$ and $m_b$ at two- and three-loop order for our NLO and NNLO predictions respectively.  We observe that the inclusion of the NNLO corrections substantially improves the overall scale dependence. This is especially true in the perturbative region specified by  $y_{\rm{cut}} >  0.01$ where we observe improvement of around a factor of two. For instance, at $y_{\rm{cut}} = 0.1$ the overall scale dependence of the jet rate at NNLO is $\{+3,-6\}$\%, compared to $\{+11,-10\}$\% for the same jet cut at NLO. Finally, we note that in the perturbative region $y_{\rm{cut}} >  0.01$ the scale variation bands provide a good estimate of the uncertainties due to the missing higher-order corrections, as the NNLO corrections lie within the NLO scale variation band. In this region we therefore expect the N$^3$LO corrections to be within the NNLO band. On the other hand, in the region $y_{\rm{cut}} <  0.01$ we observe that perturbation theory breaks down and, as expected, the scale variation bands no longer overlap. In this region the behavior of missing higher-order corrections cannot be predicted.

\begin{figure}
\begin{center}
\includegraphics[width=7cm]{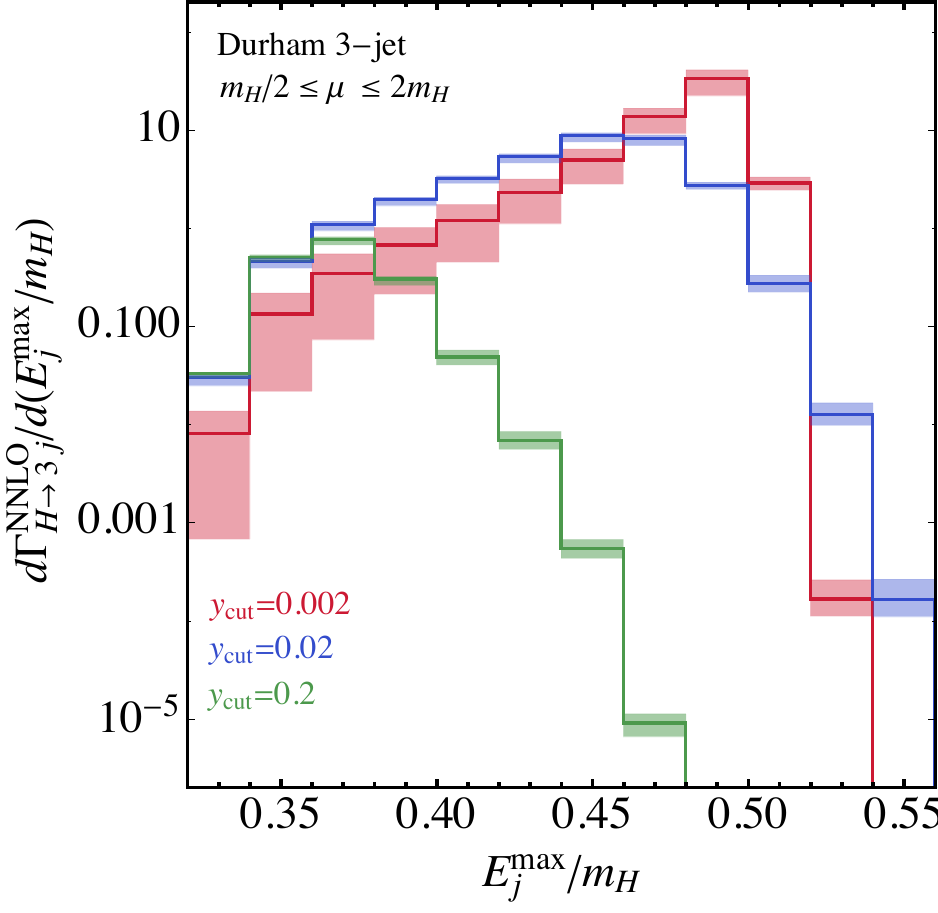}
\includegraphics[width=7cm]{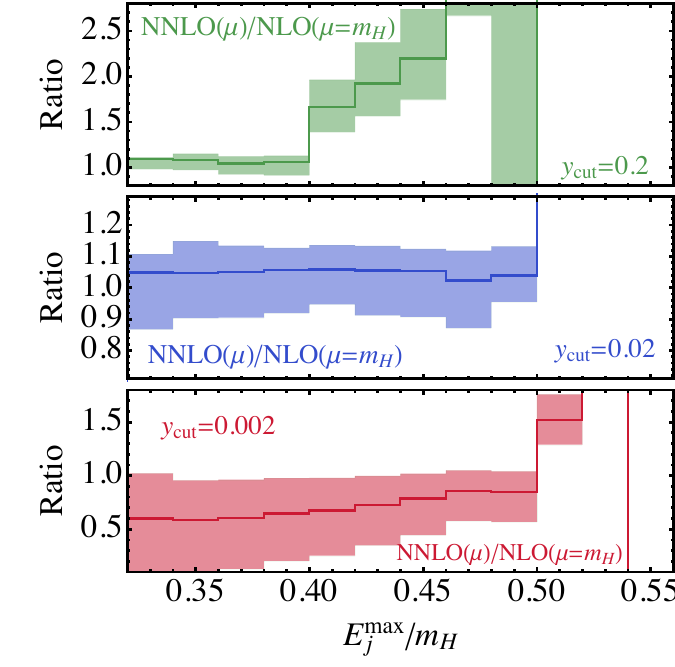}
\caption{The maximum energy of the jets (divided by the Higgs mass) for different jet-clustering options. The right-hand panel presents the ratio of the NNLO to NLO (with $\mu=m_H)$ predictions for each 
jet-clustering option. }
\label{fig:Ejmax} 
\end{center}
\end{figure}

In Fig.~\ref{fig:Ejmax} we turn our attention to differential distributions. We present the differential distribution for the energy component (rescaled by the Higgs mass) of the maximum-energy jet in three-jet events clustered with \yc$=0.2$, $0.02$, and $0.002$. Comparing the three curves we observe that as \yc decreases new phase space opens up near what would correspond to a  two-jet LO topology, which occurs around $m_H/2$. These configurations correspond to two nearly back-to-back jets with a soft/collinear third jet. In the perturbative region of \yc $= 0.2$ the prediction is more physically sensible, the majority of jets having an energy close to $m_H/3$ with the most energetic jet peaking slightly higher than this value. For the cases \yc=$0.2$ and $0.02$ the ratio of NNLO to NLO is reasonably flat and small (between $5-10\%$) until $E_{\rm{max}}/m_H$ becomes large enough that there is no LO phase space configuration possible. In this region the NLO prediction is the first non-zero 
prediction and it is hence susceptible to large corrections at the next order. The scale variation mimics that of the total jet rate and is reasonably flat in the region in which the phase space is accessible to all of the contributing parton-level phase spaces. We have also computed differential distributions for smaller values of \yc $= 2 \times 10^{-4}$. They are not presented in Fig.~\ref{fig:Ejmax} since, for such a small value of $y_{\rm{cut}}$, the differential prediction is negative over a large range of phase space. We mention these predictions here simply to note that the code can produce stable distributions with small MC uncertainties even in this region, which is relevant to the N$^3$LO results obtained in our companion paper.

\section{Conclusions}
\label{sec:conc}

In this paper we have presented the calculation of \hbbj at NNLO. We have focused on the contributions in which the Higgs boson couples directly to bottom quarks. We have performed an independent computation of the two-loop amplitudes needed at this order and found agreement with a previous calculation in the literature. Additionally, we checked our result using the known IR factorization properties of QCD when the emitted gluon becomes soft or collinear to one of the fermions and found complete agreement with the predictions in both limits. We have presented 
the two-loop amplitudes for \hbbg in full in the Appendix. 

In order to regulate the IR divergences present at this order we used the $N$-jettiness slicing technique to separate the calculation into two components. In the region of small $\tau_3$ we use SCET to construct an approximate form of the decay width. We used a computation of the 1-jettiness soft function, valid for arbitrary kinematics, coupled with the known jet functions and our computation of the hard function to construct the below-cut piece. The region  
$\tau_3 > $\tauc corresponds to the NLO computation of the $H\rightarrow b\overline{b} jj$ process, for which we calculated all of the needed helicity amplitudes using on-shell 
techniques of generalized unitarity for the one-loop pieces and BCFW recursion relations for the $H\rightarrow b\overline{b} jjj$ tree-level amplitudes. 

We implemented our results into a Monte Carlo code, based upon the existing $N$-jettiness slicing calculations of MCFM, and used it to produce differential distributions and jet rates for 
\hbbj at NNLO using the Durham jet algorithm. Our calculation neglected top quark-induced contributions, which are phenomenologically relevant. By combing our results with the available $H+j$ EFT results we can produce predictions for \hbbj relevant for the LHC and FCs which include both top and bottom Yukawa contributions. Additionally, by performing the appropriate kinematic crossings of our results we can 
compute $pp\rightarrow H+b$ at NNLO for LHC kinematics. We leave these applications to future studies. 

One of the main goals of this paper was to investigate whether a stable (slicing-independent) Monte Carlo code could be constructed for very small jet cuts. We have established this by presenting rates and differential distributions for a variety of values of the jet-clustering variable $y_{\rm{cut}}$. We are therefore able to effectively integrate out the jet at NNLO and use our results in a N$^3$LO calculation. We pursue this approach in a companion paper to this article.

\acknowledgments

We thank Ulrich Schubert for suggesting the coproduct method to simplify the 2dHPLs in the two-loop amplitudes and for helping with its implementation. We are grateful to Taushif Ahmed for clarifying the results of Ref.~\cite{Ahmed:2014pka}.  CW and RM are supported by a National Science Foundation CAREER award number PHY-1652066. Support provided by the Center for Computational Research at the University at Buffalo. 

\section*{Note added}

A previous version of this manuscript, submitted to the {\tt{arXiv}}, incorrectly claimed a disagreement with the existing two-loop calculation presented in Ref.~\cite{Ahmed:2014pka}. This was due to a misinterpretation on our part of the results of Ref.~\cite{Ahmed:2014pka}. Specifically, in two of the coefficients in Ref.~\cite{Ahmed:2014pka} the Mandelstam invariants $t$ and $u$ are retained (the remaining coefficients are functions of the variables $y$ and $z$ only), whereas our results are constructed in terms of $y$ and $z$ only. Our notation is different from Ref.~\cite{Ahmed:2014pka}  ($t \leftrightarrow u$) and the residual $t$ and $u$ dependence in the two coefficients was incorrectly evaluated in our initial comparison. Once this is altered accordingly our results are in perfect agreement. 

\appendix

\section{Formulae for renormalization and IR subtraction}
\label{renoirformulae}

The renormalization coefficients of Eqs.~\eqref{zalphaseries} and \eqref{zyseries} are defined as
\begin{align}
r_1 &= -\frac{\beta_0}{2\epsilon} \\
r_2 &= \frac{\beta_0^2}{4\epsilon^2}-\frac{\beta_1}{8\epsilon} \\
s_1 &= -\frac{3 C_F}{2\epsilon} \\
s_2 &= \frac{3}{8\epsilon^2} \left ( 3 C_F^2 + \beta_0 C_F \right ) \notag \\ &\quad -\frac{1}{8\epsilon} \left ( \frac{3}{2} C_F^2+\frac{97}{6} C_F C_A-\frac{10}{3} C_F T_R N_f \right )
\end{align}
with
\begin{align}
\beta_0 &= \frac{11}{3} C_A -\frac{4}{3} T_R N_f \\
\beta_1 &= \frac{34}{3} C_A^2 -\frac{20}{3} C_A T_R N_f -4 C_F T_R N_f
\end{align}
and $T_R = \frac{1}{2}$, $C_A=N_c$, $C_F = \frac{N_c^2-1}{2 N_c}$. 

\vspace{5mm}

The subtraction operators $I^{(\ell)}(\epsilon)$ for generic QCD processes can be found in Ref.~\cite{Catani:1998bh}. For completeness, we show here the explicit expressions for the subtraction operators in CDR for the process \hbbg:
\begin{align}
I^{(1)}(\epsilon) &= \frac{e^{\epsilon \gamma_E}}{2\,\Gamma(1-\epsilon)} \left (-\frac{m_H^2}{\mu^2}\right )^{-\epsilon} \bigg [ \left ( \frac{1}{\eps^2} + \frac{3}{2\eps} \right ) \left (C_A-2 C_F \right ) \, x^{-\eps} \notag \\ &\quad + \left ( -\frac{C_A}{\eps^2} -\frac{5 C_A}{3 \eps} + \frac{T_R N_f}{3 \eps} \right ) \left ( y^{-\eps} + z^{-\eps} \right ) \bigg ] \\
I^{(2)}(\epsilon) &=  \frac{e^{-\eps \gamma_E} \,\Gamma(1-2\eps)}{\Gamma(1-\eps)} \left (\frac{\beta_0}{2\eps} + K \right ) I^{(1)}(2\epsilon) + H^{(2)}(\eps) \notag \\ &\quad -\frac{1}{2} I^{(1)}(\epsilon) \left ( I^{(1)}(\epsilon) + \frac{\beta_0}{\eps} \right )
\end{align}
where
\begin{align}
K &= \left ( \frac{67}{18}-\zeta_2 \right ) C_A -\frac{10}{9} T_R N_f \\
H^{(2)}(\eps) &= \frac{1}{\eps} \left [ 2 H^{(2)}_q + H^{(2)}_g \right ]
\end{align}
with
\begin{align}
H^{(2)}_q &= C_F T_R N_f \left (-\frac{25}{216}+\frac{\zeta_2}{8} \right ) + C_F C_A \left (\frac{245}{864}-\frac{23 \zeta_2}{32}+\frac{13 \zeta_3}{8} \right ) \notag \\ &\quad + C_F^2 \left (-\frac{3}{32}+\frac{3 \zeta_2}{4} -\frac{3 \zeta_3}{2} \right ) \\
H^{(2)}_g &= C_A^2 \left (\frac{5}{48}+\frac{11 \zeta_2}{96} + \frac{\zeta_3}{8} \right ) + C_A T_R N_f \left (-\frac{29}{54}-\frac{\zeta_2}{24} \right ) \notag \\ &\quad + \frac{1}{4} C_F T_R N_f + \frac{5}{27} T_R^2 N_f^2 \, .
\end{align}

Finally, we present the expressions for $\mathcal{C}_0$ and $\mathcal{C}_2$ in Eqs.~\eqref{mscoeff1} and \eqref{mscoeff2}. The coefficient $\mathcal{C}_0$ corresponds to the $\eps^0$ order of the series expansion of $I^{(1)}(\epsilon)$. Explicitly:
\begin{align}
\mathcal{C}_0 &= \frac{1}{4} \left (C_A-2 C_F \right ) \left [ L(x)^2 -3 L(x) -\zeta_2 \right ] -\frac{1}{6} T_R N_f \left [ L(y)+L(z) \right ] \notag \\ &\quad +\frac{C_A}{12} \left [ 10 \left (L(y)+L(z) \right ) -3 \left (L(y)^2+L(z)^2 \right ) + 6 \zeta_2 \right ]
\end{align}
where for brevity $L(a) = \text{ln} \! \left (-\frac{m_H^2}{\mu^2}\right ) + \text{ln} \,a$. The coefficient $\mathcal{C}_2$ is defined as
\begin{align}
\mathcal{C}_2 &= \frac{1}{2} \,\mathcal{C}_0^2 + \frac{\gamma_1^{\text{cusp}}}{8} \left (\mathcal{C}_0 + \frac{3 \zeta_2}{64} \Gamma'_{0} \right ) + \frac{\beta_0}{2} \mathcal{C}'_1
\end{align}
where
\begin{align}
\gamma_1^{\text{cusp}} &= C_A \left (\frac{268}{9}-8 \zeta_2 \right ) -\frac{80}{9} T_R N_f \\
\Gamma'_{0} &= -4 \left (C_A +2 C_F \right ) \\
\mathcal{C}'_1 &= -\frac{1}{48} \left (C_A-2 C_F \right ) \left [4 L(x)^3-18 L(x)^2+6 \zeta_2 L(x)-6 \zeta_3-9 \zeta_2 \right ] \notag \\ &\quad +\frac{1}{12} T_R N_f \left [L(y)^2+L(z)^2+\zeta_2\right ] + \frac{C_A}{24} \big [ 2 L(y)^3-10 L(y)^2+3 \zeta_2 L(y) \notag \\ &\quad +2 L(z)^3-10 L(z)^2+3 \zeta_2 L(z) -6 \zeta_3-10 \zeta_2 \big ] \, .
\end{align}

\section{One-loop and two-loop coefficients for \hbbg}
\label{formulaehbbg}

We present the explicit results for the coefficients $A_m^{(0),\text{fin}}$ and $\mathcal{B}_{m;n}^{(\ell)}$ of Eq.~\eqref{bcoeffdef} as series in $N_c$. The results for the one-loop and two-loop coefficients are also available in a supplementary Mathematica-readable file attached to this paper. At tree level the coefficients are
\begin{align}
A_1^{(0),\text{fin}} = A_1^{(0)} &= 2\sqrt{2}\,\pi \left ( \frac{1}{y} + \frac{1}{z} \right ) \notag \\
A_2^{(0),\text{fin}} = A_2^{(0)} &= 2\sqrt{2}\,\pi \left ( -\frac{2}{y} \right ) \, .
\end{align}
At one loop the coefficients read:
\begin{align}
\mathcal{B}_{1;1}^{(1)} &= \frac{3}{4 N_c} + \frac{N_F}{6} - \frac{5 N_c}{3} \\
\mathcal{B}_{1;0}^{(1)} &= \frac{1}{4 N_c} \Big [ -2 H(0;y) H(1;z)+4 H(1;z) H(z;y)-2 H(0;z) H(1-z;y) \notag \\ &\quad -3 H(1-z;y)-2 H(0,1-z;y)-2 H(1-z,0;y)+4 H(z,1-z;y) \notag \\ &\quad +2 H(1,0;y)-3 H(1;z)+2 H(0,1;z)+3 \Big ] + \frac{N_F}{12} \Big [ H(0;y) + H(0;z) \Big ] \notag \\ &\quad + \frac{N_c}{12} \Big [ -6 H(0;y) H(0;z)-10 H(0;y)-6 H(1,0;y)-10 H(0;z) \notag \\ &\quad -6 H(1,0;z)-6 \zeta_2 -3 \Big ] \\
\mathcal{B}_{2;1}^{(1)} &= \mathcal{B}_{1;1}^{(1)} \\
\mathcal{B}_{2;0}^{(1)} &= \mathcal{B}_{1;0}^{(1)} -\frac{1}{4 N_c}-\frac{N_c}{4} \, .
\end{align}
At two loops:
\begin{align}
\mathcal{B}_{1;2}^{(2)} &= \frac{9}{32 N_c^2} + \frac{N_F}{4 N_c} -\frac{31}{16} +\frac{N_F^2}{24} -\frac{17 N_F N_c}{24} +\frac{35 N_c^2}{12} \\
\mathcal{B}_{1;1}^{(2)} &= -\frac{3}{16 N_c^2} \Big [ 2 H(0;y) H(1;z)+2 H(0;z) H(1-z;y)+3 H(1-z;y) \notag \\ &\quad -4 H(1;z) H(z;y)+2 H(0,1-z;y)+2 H(1-z,0;y)-4 H(z,1-z;y) \notag \\ &\quad -2 H(1,0;y)+3 H(1;z)-2 H(0,1;z)-8 \zeta_3+4 \zeta_2-3 \Big ] \notag \\ &\quad + \frac{N_F}{432 N_c} \Big [ -108 H(0;y) H(1;z)-108 H(0;z) H(1-z;y)-162 H(1-z;y) \notag \\ &\quad +216 H(1;z) H(z;y)-108 H(0,1-z;y)-108 H(1-z,0;y) \notag \\ &\quad +216 H(z,1-z;y)+27 H(0;y)+108 H(1,0;y)+27 H(0;z)-162 H(1;z) \notag \\ &\quad +108 H(0,1;z)+54\zeta_2+22 \Big ] + \frac{1}{432} \Big [ -162 H(0;y) H(0;z) \notag \\ &\quad +756 H(0;y) H(1;z)+756 H(0;z) H(1-z;y)+1134 H(1-z;y) \notag \\ &\quad -1512 H(1;z) H(z;y)+756 H(0,1-z;y)+756 H(1-z,0;y) \notag \\ &\quad -1512 H(z,1-z;y)-270 H(0;y)-918 H(1,0;y)-270 H(0;z) \notag \\ &\quad +1134 H(1;z)-756 H(0,1;z)-162 H(1,0;z)+108 \zeta_3-135 \zeta_2-97 \Big ] \notag \\ &\quad + \frac{N_F^2}{216} \Big [ 9 H(0;y)+9 H(0;z)-20 \Big ] + \frac{N_F N_c}{72} \Big [ -18 H(0;y) H(0;z) \notag \\ &\quad -51 H(0;y)-18 H(1,0;y)-51 H(0;z)-18 H(1,0;z)-24 \zeta_2+83 \Big ] \notag \\ &\quad + \frac{N_c^2}{216} \Big [ 378 H(0;y) H(0;z)+630 H(0;y)+378 H(1,0;y)+630 H(0;z) \notag \\ &\quad +378 H(1,0;z)-432 \zeta_3+477\zeta_2-721 \Big ] \\
\mathcal{B}_{1;0}^{(2)} &= \frac{1}{N_c^2} \mathcal{D}_{1;0,\text{a}}^{(2)} + \frac{N_F}{N_c} \mathcal{D}_{1;0,\text{b}}^{(2)} + \mathcal{D}_{1;0,\text{c}}^{(2)} + N_F^2\, \mathcal{D}_{1;0,\text{d}}^{(2)} + N_F N_c\, \mathcal{D}_{1;0,\text{e}}^{(2)} + N_c^2\, \mathcal{D}_{1;0,\text{f}}^{(2)} \\
\mathcal{B}_{2;2}^{(2)} &= \mathcal{B}_{1;2}^{(2)} \\
\mathcal{B}_{2;1}^{(2)} &= \mathcal{B}_{1;1}^{(2)} -\frac{3}{16 N_c^2} -\frac{N_F}{8 N_c} +\frac{11}{16} -\frac{N_F N_c}{8} +\frac{7 N_c^2}{8} \\
\mathcal{B}_{2;0}^{(2)} &= \frac{1}{N_c^2} \mathcal{D}_{2;0,\text{a}}^{(2)} + \frac{N_F}{N_c} \mathcal{D}_{2;0,\text{b}}^{(2)} + \mathcal{D}_{2;0,\text{c}}^{(2)} + N_F^2\, \mathcal{D}_{2;0,\text{d}}^{(2)} + N_F N_c\, \mathcal{D}_{2;0,\text{e}}^{(2)} + N_c^2\, \mathcal{D}_{2;0,\text{f}}^{(2)}
\end{align}
where 
\begin{align}
\mathcal{D}_{1;0,\text{a}}^{(2)} &= -\frac{11}{8} \zeta_4 +\frac{1}{32} \Big [ \,\zeta_2 \Big ( 12 H(0;y) +36 H(1;z) -16 H(1;z) H(1-z;y) \notag \\ &\quad +12 H(1-z;y) +16 H(0,1;y) -16 H(0,1-z;y) +16 H(1,1;y) \notag \\ &\quad +16 H(1-z,1;y) -32 H(1-z,1-z;y) +28  \Big ) -24 H(0;y) H(1;z) \notag \\ &\quad -18 H(1;z)-12 H(0;z) H(1-z;y)+18 H(1;z) H(1-z;y)-18 H(1-z;y) \notag \\ &\quad +40 H(1;z) H(z;y)+24 H(z;y) H(0,1;z)+24 H(0,1;z) \notag \\ &\quad +24 H(0;z) H(0,1-z;y)+24 H(1;z) H(0,1-z;y)-8 H(0,1;z) H(0,1-z;y) \notag \\ &\quad -24 H(0,1-z;y)-24 H(1;z) H(0,z;y)-12 H(0;z) H(1,0;y) \notag \\ &\quad -12 H(1;z) H(1,0;y)+8 H(0,1;z) H(1,0;y)+24 H(1,0;y) \notag \\ &\quad +12 H(0;y) H(1,0;z)-24 H(z;y) H(1,0;z) -16 H(0,1-z;y) H(1,0;z) \notag \\ &\quad -16 H(0,z;y) H(1,0;z)+24 H(0;y) H(1,1;z)-48 H(z;y) H(1,1;z) \notag \\ &\quad +16 H(0,0;y) H(1,1;z)-16 H(0,z;y) H(1,1;z)+18 H(1,1;z) \notag \\ &\quad -8 H(0,1;z) H(1,1-z;y)+12 H(0;z) H(1-z,0;y)+24 H(1;z) H(1-z,0;y) \notag \\ &\quad -8 H(0,1;z) H(1-z,0;y)-24 H(1-z,0;y)+24 H(0;z) H(1-z,1-z;y) \notag \\ &\quad +16 H(0,0;z) H(1-z,1-z;y)+32 H(0,1;z) H(1-z,1-z;y) \notag \\ &\quad +18 H(1-z,1-z;y)-24 H(1;z) H(1-z,z;y)-24 H(1;z) H(z,0;y) \notag \\ &\quad -16 H(1,1;z) H(z,0;y)-24 H(0;z) H(z,1-z;y)-48 H(1;z) H(z,1-z;y) \notag \\ &\quad +16 H(0,1;z) H(z,1-z;y)+40 H(z,1-z;y)+48 H(1;z) H(z,z;y) \notag \\ &\quad -16 H(0,1;z) H(z,z;y)+16 H(1,0;z) H(z,z;y)+32 H(1,1;z) H(z,z;y) \notag \\ &\quad -8 H(1;y) H(0,0,1;z)+16 H(1-z;y) H(0,0,1;z)+16 H(0;z) H(0,0,1-z;y) \notag \\ &\quad +16 H(1;z) H(0,0,1-z;y)-16 H(1;z) H(0,0,z;y)-8 H(1;z) H(0,1,0;y) \notag \\ &\quad +12 H(0,1,0;y)+8 H(1-z;y) H(0,1,0;z)-16 H(0;y) H(0,1,1;z) \notag \\ &\quad +16 H(z;y) H(0,1,1;z)-24 H(0,1,1;z)+16 H(1;z) H(0,1-z,0;y) \notag \\ &\quad +24 H(0,1-z,1-z;y)-8 H(1;z) H(0,1-z,z;y) \notag \\ &\quad -16 H(0;z) H(0,z,1-z;y)-16 H(1;z) H(0,z,1-z;y)-24 H(0,z,1-z;y) \notag \\ &\quad +16 H(1;z) H(0,z,z;y)-16 H(1;z) H(1,0,0;y) \notag \\ &\quad -8 H(0;y) H(1,0,1;z)-8 H(1;y) H(1,0,1;z)+24 H(1-z;y) H(1,0,1;z) \notag \\ &\quad +16 H(z;y) H(1,0,1;z)-12 H(1,0,1;z)-12 H(1,0,1-z;y) \notag \\ &\quad +8 H(1;z) H(1,0,z;y)+8 H(1-z;y) H(1,1,0;z)+12 H(1,1,0;z) \notag \\ &\quad -12 H(1,1-z,0;y)-8 H(1;z) H(1,1-z,z;y)+16 H(1;z) H(1-z,0,0;y) \notag \\ &\quad -8 H(0;z) H(1-z,0,1-z;y)+24 H(1-z,0,1-z;y) \notag \\ &\quad -8 H(1;z) H(1-z,1,0;y)-24 H(1-z,1,0;y)+24 H(1-z,1-z,0;y) \notag \\ &\quad +32 H(1;z) H(1-z,1-z,z;y)-24 H(1-z,z,1-z;y) \notag \\ &\quad -16 H(0;z) H(z,0,1-z;y)-16 H(1;z) H(z,0,1-z;y)-24 H(z,0,1-z;y) \notag \\ &\quad +16 H(1;z) H(z,0,z;y)-16 H(1;z) H(z,1-z,0;y)-24 H(z,1-z,0;y) \notag \\ &\quad -16 H(0;z) H(z,1-z,1-z;y)-48 H(z,1-z,1-z;y) \notag \\ &\quad +32 H(1;z) H(z,1-z,z;y)+16 H(1;z) H(z,z,0;y) \notag \\ &\quad +16 H(0;z) H(z,z,1-z;y)+32 H(1;z) H(z,z,1-z;y)+48 H(z,z,1-z;y) \notag \\ &\quad -32 H(1;z) H(z,z,z;y)+16 H(0,0,1,1;z)+16 H(0,0,1-z,1-z;y) \notag \\ &\quad -16 H(0,0,z,1-z;y)+16 H(0,1,0,1;z)-8 H(0,1,0,1-z;y) \notag \\ &\quad +16 H(0,1,1,0;y)+16 H(0,1,1,0;z)-8 H(0,1,1-z,0;y) \notag \\ &\quad +16 H(0,1-z,0,1-z;y)-8 H(0,1-z,1,0;y)+16 H(0,1-z,1-z,0;y) \notag \\ &\quad -8 H(0,1-z,z,1-z;y)-16 H(0,z,1-z,1-z;y)+16 H(0,z,z,1-z;y) \notag \\ &\quad +8 H(1,0,0,1;z)-16 H(1,0,0,1-z;y)+8 H(1,0,1,0;y) \notag \\ &\quad -16 H(1,0,1-z,0;y)+8 H(1,0,z,1-z;y)+16 H(1,1,0,0;y) \notag \\ &\quad +16 H(1,1,0,1;z)+16 H(1,1,1,0;y)+16 H(1,1,1,0;z) \notag \\ &\quad -16 H(1,1-z,0,0;y)-8 H(1,1-z,1,0;y)-8 H(1,1-z,z,1-z;y) \notag \\ &\quad +16 H(1-z,0,0,1-z;y)-8 H(1-z,0,1,0;y)+16 H(1-z,0,1-z,0;y) \notag \\ &\quad -16 H(1-z,1,0,0;y)-8 H(1-z,1,0,1-z;y)+16 H(1-z,1,1,0;y) \notag \\ &\quad -8 H(1-z,1,1-z,0;y)+16 H(1-z,1-z,0,0;y) \notag \\ &\quad +32 H(1-z,1-z,z,1-z;y)-16 H(z,0,1,0;y)-16 H(z,0,1-z,1-z;y) \notag \\ &\quad +16 H(z,0,z,1-z;y)-16 H(z,1-z,0,1-z;y)+16 H(z,1-z,1,0;y) \notag \\ &\quad -16 H(z,1-z,1-z,0;y)+32 H(z,1-z,z,1-z;y)+16 H(z,z,0,1-z;y) \notag \\ &\quad +16 H(z,z,1-z,0;y)+32 H(z,z,1-z,1-z;y)-32 H(z,z,z,1-z;y) \notag \\ &\quad -32 \zeta_3 H(1;y) -16 \zeta_3 H(1;z) +16 \zeta_3 H(1-z;y) -84 \zeta_3+19 \Big ] \notag \\ &\quad -\frac{z^3}{8\,x\,(1-y)^2\,(y+z)} \Big [ \zeta_2 H(0;z)-3 \zeta_3 \Big ] +\frac{z^2}{8\,x\,(1-y)\,(y+z)} \Big [ -\zeta_2 H(0;z) \notag \\ &\quad +H(1,0;z)+3 \zeta_3+\zeta_2 \Big ] +\frac{z}{8\,(1-y)\,(y+z)} \Big [ -z H(0;z) H(1-z;y) \notag \\ &\quad -2 H(0,1;z) H(1-z;y)-2 H(0;z) H(0,1-z;y)+2 H(1;z) H(0,z;y) \notag \\ &\quad -2 H(1;z) H(1-z,z;y)+2 H(0,z,1-z;y)-2 H(1-z,z,1-z;y)+H(0,1;z) \notag \\ &\quad -2 H(0,0,1;z)-2 H(1,0,1;z) \Big ] +\frac{z^2}{8\,(1-y)^2\,(y+z)} \Big [ -H(0,1;z) H(1-z;y) \notag \\ &\quad -H(0;z) H(0,1-z;y)+H(1;z) H(0,z;y)-H(1;z) H(1-z,z;y) \notag \\ &\quad +H(0,z,1-z;y)-H(1-z,z,1-z;y)-H(0,0,1;z)-H(1,0,1;z) \Big ] \notag \\ &\quad
+\frac{z}{8\,(1-z)^2\,(y+z)} \Big [ z \zeta_2 H(0;y)+3 z \zeta_2 H(1-z;y)-2 \zeta_2 H(1-z;y) \notag \\ &\quad -z H(0;y) H(0,1;z)-z H(0,1;z) H(1-z;y)-z H(1;z) H(0,z;y) \notag \\ &\quad -z H(1;z) H(1-z,z;y)+3 z H(0,1,0;y)-z H(0,z,1-z;y) \notag \\ &\quad +3 z H(1-z,1,0;y)-z H(1-z,z,1-z;y)-2 H(1-z,1,0;y) \notag \\ &\quad -2 H(0,1,0;y)+3 z \zeta_2 H(1;z)-2 \zeta_2 H(1;z)+z H(0,0,1;z)-z H(1,0,1;z) \notag \\ &\quad -3 z \zeta_3 \Big ] +\frac{1}{8\,(1-z)\,(y+z)} \Big [ -z^2 H(0;y) H(1;z)-z^2 H(0,1-z;y) \notag \\ &\quad +z^2 H(1,0;y)-z^2 H(1-z,0;y)-2 z \zeta_2 H(0;y)-2 z \zeta_2 H(1-z;y) \notag \\ &\quad +2 \zeta_2 H(1-z;y)+2 z H(0;y) H(0,1;z)+2 z H(0,1;z) H(1-z;y) \notag \\ &\quad +2 z H(1;z) H(0,z;y)+z H(1,0;y) +2 z H(1;z) H(1-z,z;y)-2 z H(0,1,0;y) \notag \\ &\quad +2 z H(0,z,1-z;y)-2 z H(1-z,1,0;y)+2 z H(1-z,z,1-z;y) \notag \\ &\quad +2 H(1-z,1,0;y)+2 H(0,1,0;y)+z^2 H(0,1;z)-2 z \zeta_2 H(1;z)+2 \zeta_2 H(1;z) \notag \\ &\quad -z H(0,1;z)-2 z H(0,0,1;z)+2 z H(1,0,1;z)+6 z \zeta_3+z \zeta_2 \Big ] \notag \\ &\quad
+\frac{1}{8\,x\,(y+z)} \Big [ -z^2 \zeta_2 H(0;y)-2 z^2 \zeta_2 H(1;y)+6 z^2 H(0;y) H(0;z) \notag \\ &\quad -z^2 H(1,0;y) H(0;z)+6 z^2 H(1,0;y)-2 z^2 H(1,1,0;y)+2 z \zeta_2 H(0;y) \notag \\ &\quad -6 z H(0;y) H(0;z)-6 z H(1,0;y)-\zeta_2 H(0;y)-z^2 \zeta_2 H(0;z)+6 z^2 H(1,0;z) \notag \\ &\quad +2 z \zeta_2 H(0;z)-7 z H(1,0;z)+4 z^2 \zeta_3+6 z^2 \zeta_2-10 z \zeta_3-7 z \zeta_2+2 \zeta_3 \Big ] \notag \\ &\quad +\frac{1}{8\,(1-z)^2} \Big [ y \zeta_2 H(1;z)+y \zeta_2 H(1-z;y)-z \zeta_2 H(0;y)-2 z \zeta_2 H(1-z;y) \notag \\ &\quad +\zeta_2 H(1-z;y)-y H(0;y) H(0,1;z)-y H(0,1;z) H(1-z;y) \notag \\ &\quad -y H(1;z) H(0,z;y)-y H(1;z) H(1-z,z;y)+y H(0,0,1;z) \notag \\ &\quad -y H(0,z,1-z;y)-y H(1,0,1;z)+y H(1-z,1,0;y) \notag \\ &\quad -y H(1-z,z,1-z;y)+z H(0;y) H(0,1;z)+z H(0,1;z) H(1-z;y) \notag \\ &\quad +z H(1;z) H(0,z;y)+z H(1;z) H(1-z,z;y)-2 z H(0,1,0;y) \notag \\ &\quad +z H(0,z,1-z;y)-2 z H(1-z,1,0;y)+H(1-z,1,0;y) \notag \\ &\quad +z H(1-z,z,1-z;y)+y \zeta_2 H(0;y)+y H(0,1,0;y)+H(0,1,0;y) \notag \\ &\quad -2 z \zeta_2 H(1;z)+\zeta_2 H(1;z)-z H(0,0,1;z)+z H(1,0,1;z)-3 y \zeta_3+3 z \zeta_3 \Big ] \notag \\ &\quad +\frac{1}{8\,(1-z)} \Big [ \zeta_2 H(1-z;y)-y H(0;y) H(1;z)+z H(0;y) H(1;z) \notag \\ &\quad +y H(1;z) H(z;y)+y H(0,1;z)-2 H(0;y) H(0,1;z)-2 H(0,1;z) H(1-z;y) \notag \\ &\quad -y H(0,1-z;y)+z H(0,1-z;y)-2 H(1;z) H(0,z;y)-z H(1,0;y) \notag \\ &\quad -y H(1-z,0;y)+z H(1-z,0;y)-2 H(1;z) H(1-z,z;y)+y H(z,1-z;y) \notag \\ &\quad -2 H(0,z,1-z;y)+H(1-z,1,0;y)-2 H(1-z,z,1-z;y)+2 \zeta_2 H(0;y) \notag \\ &\quad +y H(1,0;y)-H(1,0;y)+H(0,1,0;y)+\zeta_2 H(1;z)-z H(0,1;z)+H(0,1;z) \notag \\ &\quad +2 H(0,0,1;z)-2 H(1,0,1;z)-6 \zeta_3-\zeta_2 \Big ]
+\frac{1}{8\,(y+z)} \Big [ -4 z \zeta_2 H(0;y) \notag \\ &\quad +4 z \zeta_2 H(1;y)+2 z \zeta_2 H(1-z;y)-2 \zeta_2 H(1-z;y)+6 z H(0;y) H(0;z) \notag \\ &\quad +3 z H(0;y) H(1;z)-3 z H(0;z) H(1-z;y)+3 z H(0,1-z;y) \notag \\ &\quad +2 z H(1,0;y) H(0;z)+3 z H(1,0;y)-2 z H(0;y) H(1,0;z) \notag \\ &\quad +2 z H(1,0;z) H(1-z;y)+2 z H(0;z) H(1-z,0;y)+3 z H(1-z,0;y) \notag \\ &\quad +2 z H(0,1,0;y)+4 z H(1,1,0;y)+2 z H(1-z,1,0;y)+H(0;y) H(1,0;z) \notag \\ &\quad -2 H(1-z,1,0;y)+\zeta_2 H(0;y)-3 H(0,1,0;y)+2 z \zeta_2 H(0;z)-6 z \zeta_2 H(1;z) \notag \\ &\quad -3 z H(0,1;z)+6 z H(1,0;z)-4 z H(0,1,0;z)-6 z H(1,1,0;z)+2 H(1,1,0;z) \notag \\ &\quad +4 z \zeta_3+6 z \zeta_2-2 \zeta_3 \Big ] +\frac{z}{8\,(1-y)} \Big [ H(1;z) H(z;y)+H(z,1-z;y) \Big ] \notag \\ &\quad +\frac{1}{8\,x\,(y+z)^2} \Big [ -z^2 H(0;y) H(1,0;z)+z^2 H(0,1,0;y)+2 z H(0;y) H(1,0;z) \notag \\ &\quad -2 z H(0,1,0;y)-H(0;y) H(1,0;z)+H(0,1,0;y)-z^2 \zeta_2 H(1;z) \notag \\ &\quad +2 z^2 H(0,1,0;z)-z^2 H(1,1,0;z)+6 z \zeta_2 H(1;z)-2 \zeta_2 H(1;z) \notag \\ &\quad +2 z H(0,1,0;z)+6 z H(1,1,0;z)-2 H(1,1,0;z) \Big ] \notag \\ &\quad +\frac{1}{8\,(y+z)^2} \Big [ -2 z^2 \zeta_2 H(1-z;y)-z^2 H(0;y) H(1,0;z) \notag \\ &\quad -2 z^2 H(1,0;z) H(1-z;y)-2 z^2 H(0;z) H(1-z,0;y)+z^2 H(0,1,0;y) \notag \\ &\quad -2 z^2 H(1-z,1,0;y)-2 z H(0;y) H(1,0;z)-2 z H(1,0;z) H(1-z;y) \notag \\ &\quad +4 z H(0,1,0;y)+2 z H(1-z,1,0;y)+H(0;y) H(1,0;z)-H(0,1,0;y) \notag \\ &\quad +z^2 H(0,1,0;z)-4 z \zeta_2 H(1;z)+2 \zeta_2 H(1;z)-2 z H(0,1,0;z)-6 z H(1,1,0;z) \notag \\ &\quad +2 H(1,1,0;z) \Big ] +\frac{z^2}{8\,(1-z)\,(y+z)^2} \Big [ \zeta_2 H(1-z;y)+H(1-z,1,0;y) \notag \\ &\quad +H(0,1,0;y)+\zeta_2 H(1;z) \Big ] +\frac{(2-z)\,z}{8\,(1-z)\,(y+z)^2} \Big [ -\zeta_2 H(1-z;y) \notag \\ &\quad -H(1-z,1,0;y)-H(0,1,0;y)-\zeta_2 H(1;z) \Big ] \notag \\ &\quad +\frac{z\,(z+2)}{8\,(1-y)\,(y+z)^2} \Big [ \zeta_2 H(1-z;y)+H(1,0;z) H(1-z;y) \Big ] \notag \\ &\quad +\frac{z^2\,(1+z)}{8\,(1-y)^2\,(y+z)^2} \Big [ \zeta_2 H(1-z;y)+H(1,0;z) H(1-z;y) \Big ] \notag \\ &\quad -\frac{z^3\,(1+z)}{8\,x\,(1-y)^2\,(y+z)^2} \Big [ \zeta_2 H(1;z)+H(0,1,0;z)+H(1,1,0;z) \Big ] \notag \\ &\quad -\frac{z^2}{8\,x\,(1-y)\,(y+z)^2} \Big [ \zeta_2 H(1;z)+H(0,1,0;z)+H(1,1,0;z) \Big ] \\
\mathcal{D}_{1;0,\text{b}}^{(2)} &= \frac{z}{12\,(1-x)} \Big [ H(0;y)-H(0;z) \Big ] + \frac{1}{1296} \Big [ 270 \zeta_2 H(1-z;y) \notag \\ &\quad +279 H(0;y) H(1;z)+279 H(0;z) H(1-z;y)+324 H(1;z) H(1-z;y) \notag \\ &\quad +204 H(1-z;y)-720 H(1;z) H(z;y)-324 H(0,0;y) H(1;z) \notag \\ &\quad -324 H(0,0;z) H(1-z;y)+216 H(0;y) H(0,1;z)-216 H(0,1;z) H(1-z;y) \notag \\ &\quad -108 H(0,1;z) H(z;y)-108 H(0;z) H(0,1-z;y)+216 H(1;z) H(0,1-z;y) \notag \\ &\quad +279 H(0,1-z;y)+324 H(1;z) H(0,z;y)+54 H(1,0;y) H(0;z) \notag \\ &\quad -54 H(0;y) H(1,0;z)+216 H(1,0;z) H(1-z;y)+324 H(1,0;z) H(z;y) \notag \\ &\quad +216 H(0;y) H(1,1;z)-432 H(1,1;z) H(z;y)-108 H(0;z) H(1-z,0;y) \notag \\ &\quad +216 H(1;z) H(1-z,0;y)+279 H(1-z,0;y)+216 H(0;z) H(1-z,1-z;y) \notag \\ &\quad +324 H(1-z,1-z;y)-432 H(1;z) H(1-z,z;y) +324 H(1;z) H(z,0;y) \notag \\ &\quad +324 H(0;z) H(z,1-z;y)-432 H(1;z) H(z,1-z;y)-720 H(z,1-z;y) \notag \\ &\quad -432 H(1;z) H(z,z;y)-324 H(0,0,1-z;y)-324 H(0,1-z,0;y) \notag \\ &\quad +216 H(0,1-z,1-z;y)+324 H(0,z,1-z;y)-324 H(1-z,0,0;y) \notag \\ &\quad +216 H(1-z,0,1-z;y)+216 H(1-z,1-z,0;y)-432 H(1-z,z,1-z;y) \notag \\ &\quad +324 H(z,0,1-z;y)+324 H(z,1-z,0;y)-432 H(z,1-z,1-z;y) \notag \\ &\quad -432 H(z,z,1-z;y)-216 \zeta_2 H(1;y)+81 H(0;y)-360 H(1,0;y) \notag \\ &\quad +54 H(0,1,0;y)+324 H(1,0,0;y)-216 H(1,1,0;y)+54 \zeta_2 H(1;z) \notag \\ &\quad +189 H(0;z)+204 H(1;z)-441 H(0,1;z)-81 H(1,0;z)+324 H(1,1;z) \notag \\ &\quad -108 H(0,0,1;z)+54 H(0,1,0;z)-216 H(0,1,1;z)-216 H(1,0,1;z) \notag \\ &\quad +18 \zeta_3-54 \zeta_2-742 \Big ] \\
\mathcal{D}_{1;0,\text{c}}^{(2)} &= -\frac{5}{32} \zeta_4 + \frac{1}{1296} \Big [ \zeta_2 \Big ( 324 H(0;z) H(1;y) -756 H(1;y) +972 H(0;y) H(1;z) \notag \\ &\quad +324 H(1;y) H(1;z) +1161 H(1;z) +648 H(0;z) H(1-z;y) \notag \\ &\quad +1296 H(1;z) H(1-z;y) +1917 H(1-z;y) -1296 H(0,1;y)  \notag \\ &\quad -324 H(0,1;z) +972 H(0,1-z;y) -972 H(1,0;y) -324 H(1,0;z) \notag \\ &\quad -1296 H(1,1;y) +324 H(1,1;z) +324 H(1,1-z;y) \notag \\ &\quad +648 H(1-z,0;y) +1296 H(1-z,1-z;y) -2295 \Big ) -810 H(0;y) \notag \\ &\quad -486 H(0;y) H(0;z)-432 H(0;z)-1926 H(0;y) H(1;z)-951 H(1;z) \notag \\ &\quad -1440 H(0;z) H(1-z;y)-1782 H(1;z) H(1-z;y)-951 H(1-z;y) \notag \\ &\quad +5148 H(1;z) H(z;y)+2268 H(1;z) H(0,0;y)+2268 H(1-z;y) H(0,0;z) \notag \\ &\quad -702 H(0;y) H(0,1;z)+1188 H(1-z;y) H(0,1;z)-864 H(z;y) H(0,1;z) \notag \\ &\quad +648 H(0,0;y) H(0,1;z)+3546 H(0,1;z)+594 H(0;z) H(0,1-z;y) \notag \\ &\quad -1188 H(1;z) H(0,1-z;y)+648 H(0,0;z) H(0,1-z;y) \notag \\ &\quad -324 H(0,1;z) H(0,1-z;y)-1926 H(0,1-z;y)-1296 H(1;z) H(0,z;y) \notag \\ &\quad -648 H(0,1;z) H(0,z;y)-540 H(0;z) H(1,0;y)+486 H(1;z) H(1,0;y) \notag \\ &\quad -324 H(0,1;z) H(1,0;y)+2250 H(1,0;y)+1026 H(0;y) H(1,0;z) \notag \\ &\quad +270 H(1-z;y) H(1,0;z)-1296 H(z;y) H(1,0;z)+648 H(0,0;y) H(1,0;z) \notag \\ &\quad -1296 H(0,z;y) H(1,0;z)-324 H(1,0;y) H(1,0;z)+324 H(1,0;z) \notag \\ &\quad -1188 H(0;y) H(1,1;z)+2376 H(z;y) H(1,1;z)
+648 H(0,z;y) H(1,1;z) \notag \\ &\quad -1782 H(1,1;z)-324 H(0,1;z) H(1,1-z;y)+324 H(1,0;z) H(1,1-z;y) \notag \\ &\quad +1566 H(0;z) H(1-z,0;y)-1188 H(1;z) H(1-z,0;y) \notag \\ &\quad +648 H(0,0;z) H(1-z,0;y)+972 H(0,1;z) H(1-z,0;y) \notag \\ &\quad +972 H(1,0;z) H(1-z,0;y)-1926 H(1-z,0;y) \notag \\ &\quad -1188 H(0;z) H(1-z,1-z;y)+1296 H(1,0;z) H(1-z,1-z;y) \notag \\ &\quad -1782 H(1-z,1-z;y)+2376 H(1;z) H(1-z,z;y) \notag \\ &\quad -1944 H(0,1;z) H(1-z,z;y)+648 H(1,0;z) H(1-z,z;y) \notag \\ &\quad -1296 H(1;z) H(z,0;y)+648 H(0,1;z) H(z,0;y)+648 H(1,1;z) H(z,0;y) \notag \\ &\quad -1296 H(0;z) H(z,1-z;y)+2376 H(1;z) H(z,1-z;y) \notag \\ &\quad -648 H(0,1;z) H(z,1-z;y)+648 H(1,0;z) H(z,1-z;y) \notag \\ &\quad +5148 H(z,1-z;y)+432 H(1;z) H(z,z;y)-3240 H(0,1;z) H(z,z;y) \notag \\ &\quad +648 H(1,0;z) H(z,z;y)-1296 H(1,1;z) H(z,z;y)-648 H(0;y) H(0,0,1;z) \notag \\ &\quad -324 H(1;y) H(0,0,1;z)-1296 H(1-z;y) H(0,0,1;z) \notag \\ &\quad -2592 H(z;y) H(0,0,1;z)+108 H(0,0,1;z)+2268 H(0,0,1-z;y) \notag \\ &\quad +648 H(1;z) H(0,0,z;y)-324 H(0;z) H(0,1,0;y)+324 H(1;z) H(0,1,0;y) \notag \\ &\quad -2484 H(0,1,0;y)-324 H(0;y) H(0,1,0;z)+324 H(1;y) H(0,1,0;z) \notag \\ &\quad +1296 H(1-z;y) H(0,1,0;z)+648 H(z;y) H(0,1,0;z)
-1512 H(0,1,0;z) \notag \\ &\quad -648 H(z;y) H(0,1,1;z)+1188 H(0,1,1;z)+2268 H(0,1-z,0;y) \notag \\ &\quad -648 H(0;z) H(0,1-z,1-z;y)-1188 H(0,1-z,1-z;y) \notag \\ &\quad +324 H(1;z) H(0,1-z,z;y)+1296 H(1;z) H(0,z,0;y) \notag \\ &\quad -1296 H(0;z) H(0,z,1-z;y)+648 H(1;z) H(0,z,1-z;y) \notag \\ &\quad -1296 H(0,z,1-z;y)+648 H(1;z) H(0,z,z;y)-648 H(0;z) H(1,0,0;y) \notag \\ &\quad +648 H(1;z) H(1,0,0;y)-2268 H(1,0,0;y)+648 H(1-z;y) H(1,0,0;z) \notag \\ &\quad +324 H(0;y) H(1,0,1;z)-324 H(1;y) H(1,0,1;z) \notag \\ &\quad +972 H(1-z;y) H(1,0,1;z)-648 H(z;y) H(1,0,1;z)+1674 H(1,0,1;z) \notag \\ &\quad -648 H(0;z) H(1,0,1-z;y)+486 H(1,0,1-z;y)+324 H(1;z) H(1,0,z;y) \notag \\ &\quad -756 H(1,1,0;y)+648 H(0;y) H(1,1,0;z)+324 H(1;y) H(1,1,0;z) \notag \\ &\quad +2268 H(1-z;y) H(1,1,0;z)+648 H(z;y) H(1,1,0;z) \notag \\ &\quad -324 H(0;z) H(1,1-z,0;y)+486 H(1,1-z,0;y) \notag \\ &\quad -324 H(1;z) H(1,1-z,z;y)+648 H(0;z) H(1-z,0,0;y) \notag \\ &\quad +2268 H(1-z,0,0;y)+324 H(0;z) H(1-z,0,1-z;y) \notag \\ &\quad -1188 H(1-z,0,1-z;y)+648 H(1;z) H(1-z,0,z;y) \notag \\ &\quad +324 H(0;z) H(1-z,1,0;y)+972 H(1;z) H(1-z,1,0;y) \notag \\ &\quad +1458 H(1-z,1,0;y)+648 H(0;z) H(1-z,1-z,0;y) \notag \\ &\quad -1188 H(1-z,1-z,0;y)+648 H(1;z) H(1-z,z,0;y)
\notag \\ &\quad +648 H(0;z) H(1-z,z,1-z;y)+2376 H(1-z,z,1-z;y) \notag \\ &\quad -2592 H(1;z) H(1-z,z,z;y)+648 H(1;z) H(z,0,1-z;y) \notag \\ &\quad -1296 H(z,0,1-z;y)+648 H(1;z) H(z,0,z;y)+648 H(1;z) H(z,1-z,0;y) \notag \\ &\quad -1296 H(z,1-z,0;y)+648 H(0;z) H(z,1-z,1-z;y) \notag \\ &\quad +2376 H(z,1-z,1-z;y)-1296 H(1;z) H(z,1-z,z;y) \notag \\ &\quad +648 H(1;z) H(z,z,0;y)+648 H(0;z) H(z,z,1-z;y) \notag \\ &\quad -1296 H(1;z) H(z,z,1-z;y)+432 H(z,z,1-z;y) \notag \\ &\quad -3888 H(1;z) H(z,z,z;y)-648 H(0,0,1,0;y)+1296 H(0,0,1,0;z) \notag \\ &\quad +648 H(0,0,z,1-z;y)+324 H(0,1,0,1-z;y)-1296 H(0,1,1,0;y) \notag \\ &\quad +648 H(0,1,1,0;z)+324 H(0,1,1-z,0;y)+972 H(0,1-z,1,0;y) \notag \\ &\quad +324 H(0,1-z,z,1-z;y)+1296 H(0,z,0,1-z;y)+1296 H(0,z,1-z,0;y) \notag \\ &\quad +648 H(0,z,1-z,1-z;y)+648 H(0,z,z,1-z;y)-324 H(1,0,0,1;z) \notag \\ &\quad +648 H(1,0,0,1-z;y)-1296 H(1,0,1,0;y)+648 H(1,0,1,0;z) \notag \\ &\quad +648 H(1,0,1-z,0;y)+324 H(1,0,z,1-z;y)-1296 H(1,1,0,0;y) \notag \\ &\quad +648 H(1,1,0,1;z)-1296 H(1,1,1,0;y)+972 H(1,1,1,0;z) \notag \\ &\quad +648 H(1,1-z,0,0;y)-324 H(1,1-z,z,1-z;y)+648 H(1-z,0,1,0;y) \notag \\ &\quad +648 H(1-z,0,z,1-z;y)+648 H(1-z,1,0,0;y) \notag \\ &\quad +972 H(1-z,1,0,1-z;y)+972 H(1-z,1,1-z,0;y) \notag \\ &\quad +1296 H(1-z,1-z,1,0;y)+648 H(1-z,z,0,1-z;y) \notag \\ &\quad +648 H(1-z,z,1-z,0;y)-2592 H(1-z,z,z,1-z;y) \notag \\ &\quad +648 H(z,0,1-z,1-z;y)+648 H(z,0,z,1-z;y) \notag \\ &\quad +648 H(z,1-z,0,1-z;y)+648 H(z,1-z,1-z,0;y) \notag \\ &\quad -1296 H(z,1-z,z,1-z;y)+648 H(z,z,0,1-z;y)+648 H(z,z,1-z,0;y) \notag \\ &\quad -1296 H(z,z,1-z,1-z;y)-3888 H(z,z,z,1-z;y)-972 \zeta_3 H(1;y) \notag \\ &\quad +648 \zeta_3 H(1;z) +1620 \zeta_3 H(1-z;y) +1845 \zeta_3+2464 \Big ] +\frac{3 \zeta_3 \,z^3}{8\,x\,(1-y)^2\,(y+z)} \notag \\ &\quad + \frac{z^2}{8\,x\,(1-y)\,(y+z)} \Big [\zeta_2+H(1,0;z)+3\zeta_3 \Big ] + \frac{z}{8\,(1-y)\,(y+z)} \Big [ 2 \zeta_2 H(1-z;y) \notag \\ &\quad -z H(0;z) H(1-z;y)-2 H(0,1;z) H(1-z;y)-2 H(0;z) H(0,1-z;y) \notag \\ &\quad +2 H(1;z) H(0,z;y)+2 H(1,0;z) H(1-z;y)-2 H(1;z) H(1-z,z;y) \notag \\ &\quad +2 H(0,z,1-z;y)-2 H(1-z,z,1-z;y)-6 H(0;y)+2 \zeta_2 H(0;z) \notag \\ &\quad +2 \zeta_2 H(1;z)+H(0,1;z)-2 H(0,0,1;z)+2 H(0,1,0;z)-2 H(1,0,1;z) \notag \\ &\quad +2 H(1,1,0;z) \Big ] + \frac{z^2}{8\,(1-y)^2\,(y+z)} \Big [ \zeta_2 H(1-z;y)-H(0;z) H(0,1-z;y) \notag \\ &\quad -H(0,1;z) H(1-z;y)+H(1;z) H(0,z;y)+H(1,0;z) H(1-z;y) \notag \\ &\quad -H(1;z) H(1-z,z;y)+H(0,z,1-z;y)-H(1-z,z,1-z;y) \notag \\ &\quad +\zeta_2 H(0;z)+\zeta_2 H(1;z)-H(0,0,1;z)+H(0,1,0;z)-H(1,0,1;z) \notag \\ &\quad +H(1,1,0;z) \Big ] + \frac{z^2}{8\,(1-z)^2\,(y+z)} \Big [ \zeta_2 H(1-z;y)-H(0;y) H(0,1;z) \notag \\ &\quad -H(0,1;z) H(1-z;y)-H(1;z) H(0,z;y)-H(1;z) H(1-z,z;y) \notag \\ &\quad -H(0,z,1-z;y)+H(1-z,1,0;y)-H(1-z,z,1-z;y)+\zeta_2 H(0;y) \notag \\ &\quad +H(0,1,0;y)+\zeta_2 H(1;z)+H(0,0,1;z)-H(1,0,1;z)-3 \zeta_3 \Big ] \notag \\ &\quad + \frac{z}{8\,(1-z)\,(y+z)} \Big [ -2 \zeta_2 H(1-z;y)-z H(0;y) H(1;z)+2 H(0;y) H(0,1;z) \notag \\ &\quad +2 H(0,1;z) H(1-z;y)-z H(0,1-z;y)+2 H(1;z) H(0,z;y)+z H(1,0;y) \notag \\ &\quad -z H(1-z,0;y)+2 H(1;z) H(1-z,z;y)+2 H(0,z,1-z;y) \notag \\ &\quad -2 H(1-z,1,0;y)+2 H(1-z,z,1-z;y)-2 \zeta_2 H(0;y)+H(1,0;y) \notag \\ &\quad -2 H(0,1,0;y)-2 \zeta_2 H(1;z)+6 H(0;z)+z H(0,1;z)-H(0,1;z) \notag \\ &\quad -2 H(0,0,1;z)+2 H(1,0,1;z)+6 \zeta_3+\zeta_2 \Big ] + \frac{z}{8\,x\,(y+z)} \Big [ 6 z H(0;y) H(0;z) \notag \\ &\quad -6 H(0;y) H(0;z)+6 z H(1,0;y)-6 H(1,0;y)+6 z H(1,0;z)-7 H(1,0;z) \notag \\ &\quad +6 z \zeta_2-6 \zeta_3-7 \zeta_2 \Big ] + \frac{1}{8\,(1-z)^2} \Big [ y \zeta_2 H(1;z)+y \zeta_2 H(1-z;y)-z \zeta_2 H(0;y) \notag \\ &\quad -z \zeta_2 H(1-z;y)-y H(0;y) H(0,1;z)-y H(0,1;z) H(1-z;y) \notag \\ &\quad -y H(1;z) H(0,z;y)-y H(1;z) H(1-z,z;y)+y H(0,0,1;z) \notag \\ &\quad -y H(0,z,1-z;y)-y H(1,0,1;z)+y H(1-z,1,0;y) \notag \\ &\quad -y H(1-z,z,1-z;y)+z H(0;y) H(0,1;z)+z H(0,1;z) H(1-z;y) \notag \\ &\quad +z H(1;z) H(0,z;y)+z H(1;z) H(1-z,z;y)-z H(0,1,0;y) \notag \\ &\quad +z H(0,z,1-z;y)-z H(1-z,1,0;y)+z H(1-z,z,1-z;y) \notag \\ &\quad +y \zeta_2 H(0;y)+y H(0,1,0;y)-z \zeta_2 H(1;z)-z H(0,0,1;z)+z H(1,0,1;z) \notag \\ &\quad -3 y \zeta_3+3 z \zeta_3 \Big ] + \frac{1}{8\,(1-z)} \Big [ 2 \zeta_2 H(1-z;y)-y H(0;y) H(1;z) \notag \\ &\quad +z H(0;y) H(1;z)+y H(1;z) H(z;y)+y H(0,1;z)-2 H(0;y) H(0,1;z) \notag \\ &\quad -2 H(0,1;z) H(1-z;y)-y H(0,1-z;y)+z H(0,1-z;y) \notag \\ &\quad -2 H(1;z) H(0,z;y)-z H(1,0;y)-y H(1-z,0;y)+z H(1-z,0;y) \notag \\ &\quad -2 H(1;z) H(1-z,z;y)+y H(z,1-z;y)-2 H(0,z,1-z;y) \notag \\ &\quad +2 H(1-z,1,0;y)-2 H(1-z,z,1-z;y)+2 \zeta_2 H(0;y)+y H(1,0;y) \notag \\ &\quad -H(1,0;y)+2 H(0,1,0;y)+2 \zeta_2 H(1;z)-6 H(0;z)-z H(0,1;z) \notag \\ &\quad +H(0,1;z)+2 H(0,0,1;z)-2 H(1,0,1;z)-6 \zeta_3-\zeta_2 \Big ] \notag \\ &\quad + \frac{1}{24\,(y+z)} \Big [ -72 \zeta_2 H(1-z;y)+7 z H(0;y)+18 z H(0;y) H(0;z) \notag \\ &\quad +9 z H(0;y) H(1;z)-9 z H(0;z) H(1-z;y)+36 H(0,1;z) H(1-z;y) \notag \\ &\quad +9 z H(0,1-z;y)+9 z H(1,0;y)-18 H(1,0;y) H(1;z) \notag \\ &\quad -18 H(1,0;z) H(1-z;y)+9 z H(1-z,0;y)+36 H(1;z) H(1-z,z;y) \notag \\ &\quad -18 H(1,0,1-z;y)-18 H(1,1-z,0;y)-18 H(1-z,1,0;y) \notag \\ &\quad +36 H(1-z,z,1-z;y)+36 \zeta_2 H(1;y)+36 H(1,1,0;y)-36 \zeta_2 H(1;z) \notag \\ &\quad -7 z H(0;z)-9 z H(0,1;z)+18 z H(1,0;z)+18 H(1,0,1;z)+18 z \zeta_2 \Big ] \notag \\ &\quad + \frac{z}{8\,(1-y)} \Big [ H(1;z) H(z;y) + H(z,1-z;y) \Big ] \\
\mathcal{D}_{1;0,\text{d}}^{(2)} &= \frac{1}{432} \Big [ 3 H(0;y) H(0;z)-20 H(0;y)+15 H(0,0;y)-20 H(0;z) \notag \\ &\quad +15 H(0,0;z)+6 \zeta_2 \Big ] \\
\mathcal{D}_{1;0,\text{e}}^{(2)} &= -\frac{19}{36} \zeta_3 +\frac{1}{12\,(1-x)\,x} \Big [ y^2 H(0;y) H(0;z) \notag \\ &\quad +y^2 H(1,0;z)+y^2 H(1,0;y)+2 y H(0;z)-2 y H(0;y) \Big ] \notag \\ &\quad + \frac{1}{1296\,x} \Big [ -108 y H(0;y) H(0;z)+180 H(0;y) H(0;z)-324 H(0;y) H(0,0;z) \notag \\ &\quad -270 H(0;y) H(1,0;z)-216 y H(0;z)-324 H(0,0;y) H(0;z) \notag \\ &\quad -54 H(1,0;y) H(0;z)-108 y H(1,0;z)-216 H(1,0;z) H(1-z;y) \notag \\ &\quad -216 H(0;z) H(1-z,0;y)-216 H(1-z,1,0;y)+216 y H(0;y) \notag \\ &\quad +855 H(0;y)-738 H(0,0;y)-108 y H(1,0;y)+360 H(1,0;y) \notag \\ &\quad -54 H(0,1,0;y)-324 H(1,0,0;y)+216 H(1,1,0;y)+639 H(0;z) \notag \\ &\quad -738 H(0,0;z)+360 H(1,0;z)-54 H(0,1,0;z)-324 H(1,0,0;z) \notag \\ &\quad -136 \Big ] -\frac{1-x}{1296\,x} \Big [ 180 H(0;y) H(0;z)-324 H(0;y) H(0,0;z) \notag \\ &\quad -270 H(0;y) H(1,0;z)-324 H(0,0;y) H(0;z)-54 H(1,0;y) H(0;z) \notag \\ &\quad -216 H(1,0;z) H(1-z;y)-216 H(0;z) H(1-z,0;y) \notag \\ &\quad -216 H(1-z,1,0;y)+855 H(0;y)-738 H(0,0;y)+360 H(1,0;y) \notag \\ &\quad -54 H(0,1,0;y)-324 H(1,0,0;y)+216 H(1,1,0;y)+639 H(0;z) \notag \\ &\quad -738 H(0,0;z)+360 H(1,0;z)-54 H(0,1,0;z)-324 H(1,0,0;z) \notag \\ &\quad -136 \Big ] -\frac{y z \,\zeta_2}{12\,(1-x)\,x} -\frac{1}{144} \zeta_2 \Big [ 24 H(1-z;y)+12 H(0;y)-24 H(1;y) \notag \\ &\quad +12 H(0;z)+1 \Big ] \\
\mathcal{D}_{1;0,\text{f}}^{(2)} &= \frac{11}{4} \zeta_4 +\frac{1}{2592} \Big [ \zeta_2 \Big ( 3618 H(0;y) +1944 H(0;y) H(0;z) +2646 H(0;z) \notag \\ &\quad +648 H(0;z) H(1;y) -2376 H(1;y) +648 H(1;y) H(1;z) +2916 H(1;z) \notag \\ &\quad -648 H(0;z) H(1-z;y) -1296 H(1;z) H(1-z;y) +3348 H(1-z;y) \notag \\ &\quad -1296 H(1;z) H(z;y) +1296 H(0,1;y) +1296 H(0,1;z) +1296 H(0,1-z;y) \notag \\ &\quad +1944 H(1,0;y) +1296 H(1,0;z) +1296 H(1,1;y) +648 H(1,1;z) \notag \\ &\quad +648 H(1,1-z;y) -648 H(1-z,0;y) -1296 H(1-z,1;y) \notag \\ &\quad -1296 H(z,1-z;y) +954 \Big ) -2652 H(0;y)-2700 H(0;y) H(0;z) \notag \\ &\quad -6000 H(0;z)+4536 H(0;z) H(0,0;y)+5760 H(0,0;y) \notag \\ &\quad +4536 H(0;y) H(0,0;z)+1296 H(0,0;y) H(0,0;z)+5760 H(0,0;z) \notag \\ &\quad +108 H(0;z) H(1,0;y)-4500 H(1,0;y)+4428 H(0;y) H(1,0;z) \notag \\ &\quad +3348 H(1-z;y) H(1,0;z)+1296 H(0,1-z;y) H(1,0;z) \notag \\ &\quad +648 H(1,0;y) H(1,0;z)-4500 H(1,0;z)+648 H(1,0;z) H(1,1-z;y) \notag \\ &\quad +3348 H(0;z) H(1-z,0;y)-648 H(1,0;z) H(1-z,0;y) \notag \\ &\quad -1296 H(1,0;z) H(z,0;y)-1296 H(1,0;z) H(z,1-z;y) \notag \\ &\quad +648 H(0;z) H(0,1,0;y)+2052 H(0,1,0;y)+648 H(0;y) H(0,1,0;z) \notag \\ &\quad +648 H(1;y) H(0,1,0;z)+648 H(1-z;y) H(0,1,0;z) \notag \\ &\quad +1296 H(z;y) H(0,1,0;z)+3024 H(0,1,0;z)+1296 H(0;z) H(0,1-z,0;y) \notag \\ &\quad +1296 H(0;z) H(1,0,0;y)+4536 H(1,0,0;y)+1296 H(0;y) H(1,0,0;z) \notag \\ &\quad +4536 H(1,0,0;z)-2376 H(1,1,0;y)+648 H(1;y) H(1,1,0;z) \notag \\ &\quad -1296 H(1-z;y) H(1,1,0;z)-1296 H(z;y) H(1,1,0;z)+2916 H(1,1,0;z) \notag \\ &\quad +648 H(0;z) H(1,1-z,0;y)-648 H(0;z) H(1-z,1,0;y) \notag \\ &\quad +3348 H(1-z,1,0;y)-1296 H(0;z) H(z,1-z,0;y)+1296 H(0,0,1,0;y) \notag \\ &\quad +2592 H(0,0,1,0;z)+1296 H(0,1,1,0;y)+1296 H(0,1,1,0;z) \notag \\ &\quad +1296 H(0,1-z,1,0;y)+1944 H(1,0,1,0;y)+2592 H(1,0,1,0;z) \notag \\ &\quad +1296 H(1,1,0,0;y)+1296 H(1,1,0,0;z)+1296 H(1,1,1,0;y) \notag \\ &\quad +648 H(1,1,1,0;z)+648 H(1,1-z,1,0;y)+648 H(1-z,0,1,0;y) \notag \\ &\quad -1296 H(1-z,1,1,0;y)+1296 H(z,0,1,0;y)-1296 H(z,1-z,1,0;y) \notag \\ &\quad +\zeta_3 \Big ( -648 H(0;y) -648 H(0;z) +648 H(1;y) +3240 H(1;z) \notag \\ &\quad +2592 H(1-z;y) +6552 \Big ) -379 \Big ] -\frac{3\,z^2}{4\,x\,(1-y)\,(y+z)} H(0;y) \notag \\ &\quad -\frac{3\,z}{4\,(1-z)\,(y+z)} H(0;z) +\frac{1}{24\,x\,(y+z)} \Big [ 3 z^2 \zeta_2 H(0;y)+6 z^2 \zeta_2 H(1;y) \notag \\ &\quad -38 z^2 H(0;y) H(0;z)+3 z^2 H(1,0;y) H(0;z)-38 z^2 H(1,0;y) \notag \\ &\quad +6 z^2 H(1,1,0;y)-6 z \zeta_2 H(0;y)+18 z H(0;y) +38 z H(0;y) H(0;z) \notag \\ &\quad +38 z H(1,0;y)+3 \zeta_2 H(0;y)+3 z^2 \zeta_2 H(0;z)-38 z^2 H(1,0;z)+38 z H(1,0;z) \notag \\ &\quad -12 z^2 \zeta_3-38 z^2 \zeta_2+12 z \zeta_3+38 z \zeta_2-6 \zeta_3 \Big ] +\frac{3}{4\,(1-z)} H(0;z) \notag \\ &\quad
+\frac{1}{24\,(y+z)} \Big [ -6 z \zeta_2 H(0;y)+24 z \zeta_2 H(1;y)-6 z \zeta_2 H(1-z;y)-31 z H(0;y) \notag \\ &\quad -38 z H(0;y) H(0;z)+12 z H(1,0;y) H(0;z)-38 z H(1,0;y) \notag \\ &\quad -12 z H(0;y) H(1,0;z)-6 z H(1,0;z) H(1-z;y)-6 z H(0;z) H(1-z,0;y) \notag \\ &\quad +12 z H(0,1,0;y)+24 z H(1,1,0;y)-6 z H(1-z,1,0;y)-3 H(0;y) H(1,0;z) \notag \\ &\quad -3 \zeta_2 H(0;y)+3 H(0,1,0;y)+12 z \zeta_2 H(0;z)-18 z \zeta_2 H(1;z)-6 \zeta_2 H(1;z) \notag \\ &\quad +31 z H(0;z)-38 z H(1,0;z)-6 z H(0,1,0;z)-18 z H(1,1,0;z) \notag \\ &\quad -6 H(1,1,0;z)-12 z \zeta_3-38 z \zeta_2+6 \zeta_3 \Big ] +\frac{1}{8\,x\,(y+z)^2} \Big [ z^2 H(0;y) H(1,0;z) \notag \\ &\quad -z^2 H(0,1,0;y)-2 z H(0;y) H(1,0;z)+2 z H(0,1,0;y)+H(0;y) H(1,0;z) \notag \\ &\quad -H(0,1,0;y)+2 z^2 \zeta_2 H(1;z)-z^2 H(0,1,0;z)+2 z^2 H(1,1,0;z) \notag \\ &\quad -4 z \zeta_2 H(1;z)+2 \zeta_2 H(1;z)-4 z H(1,1,0;z)+2 H(1,1,0;z) \Big ] \notag \\ &\quad +\frac{1}{8\,(y+z)^2} \Big [ 2 z^2 \zeta_2 H(1-z;y)+z^2 H(0;y) H(1,0;z) \notag \\ &\quad +2 z^2 H(1,0;z) H(1-z;y)+2 z^2 H(0;z) H(1-z,0;y)-z^2 H(0,1,0;y) \notag \\ &\quad +2 z^2 H(1-z,1,0;y)+2 z H(0;y) H(1,0;z)-2 z H(0,1,0;y) \notag \\ &\quad -H(0;y) H(1,0;z)+H(0,1,0;y)-z^2 H(0,1,0;z)+4 z \zeta_2 H(1;z) \notag \\ &\quad -2 \zeta_2 H(1;z)+4 z H(1,1,0;z)-2 H(1,1,0;z) \Big ]
\end{align}
and
\begin{align}
\mathcal{D}_{2;0,\text{a}}^{(2)} &= \mathcal{D}_{1;0,\text{a}}^{(2)} +\frac{1}{32} \Big [ 4 H(0;y) H(1;z)+4 H(0;z) H(1-z;y)-18 H(1-z;y) \notag \\ &\quad -24 H(1;z) H(z;y)+4 H(0,1-z;y)+4 H(1-z,0;y)-24 H(z,1-z;y) \notag \\ &\quad -24 H(0;y)-4 H(1,0;y)-18 H(1;z)-4 H(0,1;z)-7 \Big ] \notag \\ &\quad +\frac{z}{8\,(1-y)\,(y+z)} \Big [ -2 \zeta_2 H(1-z;y)+4 z H(0;z) H(1-z;y) \notag \\ &\quad +2 H(0,1;z) H(1-z;y)-2 H(1,0;z) H(1-z;y)+2 H(0;z) H(0,1-z;y) \notag \\ &\quad -2 H(1;z) H(0,z;y)+2 H(1;z) H(1-z,z;y)-2 H(0,z,1-z;y) \notag \\ &\quad +2 H(1-z,z,1-z;y)-H(0,1;z)+2 H(0,0,1;z)+2 H(1,0,1;z) \Big ] \notag \\ &\quad +\frac{z^2}{8\,(1-y)^2\,(y+z)} \Big [ -\zeta_2 H(1-z;y)+H(0,1;z) H(1-z;y) \notag \\ &\quad -H(1,0;z) H(1-z;y)+H(0;z) H(0,1-z;y)-H(1;z) H(0,z;y) \notag \\ &\quad +H(1;z) H(1-z,z;y)-H(0,z,1-z;y)+H(1-z,z,1-z;y) \notag \\ &\quad +H(0,0,1;z)+H(1,0,1;z) \Big ] +\frac{z^2}{8\,(1-z)^2\,(y+z)} \Big [ -\zeta_2 H(1-z;y) \notag \\ &\quad +H(0;y) H(0,1;z)+H(0,1;z) H(1-z;y)+H(1;z) H(0,z;y) \notag \\ &\quad +H(1;z) H(1-z,z;y)+H(0,z,1-z;y)-H(1-z,1,0;y) \notag \\ &\quad +H(1-z,z,1-z;y)-\zeta_2 H(0;y)-H(0,1,0;y)-\zeta_2 H(1;z)-H(0,0,1;z) \notag \\ &\quad +H(1,0,1;z)+3 \zeta_3 \Big ] +\frac{z}{8\,(1-z)\,(y+z)} \Big [ 2 \zeta_2 H(1-z;y)+4 z H(0;y) H(1;z) \notag \\ &\quad -3 H(0;y) H(1;z)-2 H(0;y) H(0,1;z)-2 H(0,1;z) H(1-z;y) \notag \\ &\quad +4 z H(0,1-z;y)-3 H(0,1-z;y)-2 H(1;z) H(0,z;y)-4 z H(1,0;y) \notag \\ &\quad +4 z H(1-z,0;y)-3 H(1-z,0;y)-2 H(1;z) H(1-z,z;y) \notag \\ &\quad -2 H(0,z,1-z;y)+2 H(1-z,1,0;y)-2 H(1-z,z,1-z;y)+2 \zeta_2 H(0;y) \notag \\ &\quad +2 H(1,0;y)+2 H(0,1,0;y)+2 \zeta_2 H(1;z)-4 z H(0,1;z)+4 H(0,1;z) \notag \\ &\quad +2 H(0,0,1;z)-2 H(1,0,1;z)-6 \zeta_3-\zeta_2 \Big ]
+\frac{1}{8\,x\,(y+z)} \Big [ 2 z^2 H(0;y) H(0;z) \notag \\ &\quad +2 z^2 H(1,0;y)+4 z \zeta_2 H(0;y)-4 z \zeta_2 H(1;y)-6 z H(0;y)-8 z H(0;y) H(0;z) \notag \\ &\quad -2 z H(1,0;y) H(0;z)-6 z H(1,0;y)+4 z H(0;y) H(1,0;z)-4 z H(0,1,0;y) \notag \\ &\quad -4 z H(1,1,0;y)-4 H(0;y) H(1,0;z)-4 \zeta_2 H(0;y)+6 H(0;y)-H(1,0;y) \notag \\ &\quad +4 H(0,1,0;y)+2 z^2 H(1,0;z)-4 z \zeta_2 H(0;z) +6 z \zeta_2 H(1;z) \notag \\ &\quad -8 \zeta_2 H(1;z)+6 z H(0;z)-9 z H(1,0;z)+6 z H(1,1,0;z) +H(1,0;z) \notag \\ &\quad -8 H(1,1,0;z)+2 z^2 \zeta_2+2 z \zeta_3-7 z \zeta_2+8 \zeta_3+\zeta_2 \Big ] \notag \\ &\quad +\frac{1}{8\,(1-z)^2} \Big [ -y \zeta_2 H(1;z)-y \zeta_2 H(1-z;y)+z \zeta_2 H(0;y)+z \zeta_2 H(1-z;y) \notag \\ &\quad +y H(0;y) H(0,1;z)+y H(0,1;z) H(1-z;y)+y H(1;z) H(0,z;y) \notag \\ &\quad +y H(1;z) H(1-z,z;y)-y H(0,0,1;z)+y H(0,z,1-z;y)+y H(1,0,1;z) \notag \\ &\quad -y H(1-z,1,0;y)+y H(1-z,z,1-z;y)-z H(0;y) H(0,1;z) \notag \\ &\quad -z H(0,1;z) H(1-z;y)-z H(1;z) H(0,z;y)-z H(1;z) H(1-z,z;y) \notag \\ &\quad +z H(0,1,0;y)-z H(0,z,1-z;y)+z H(1-z,1,0;y) \notag \\ &\quad -z H(1-z,z,1-z;y)+y (-\zeta_2) H(0;y)-y H(0,1,0;y)+z \zeta_2 H(1;z) \notag \\ &\quad +z H(0,0,1;z)-z H(1,0,1;z)+3 y \zeta_3-3 z \zeta_3 \Big ] \notag \\ &\quad +\frac{1}{8\,(1-z)} \Big [ -2 \zeta_2 H(1-z;y)+4 y H(0;y) H(1;z)-4 z H(0;y) H(1;z) \notag \\ &\quad +3 H(0;y) H(1;z)-4 y H(1;z) H(z;y)-4 y H(0,1;z)+2 H(0;y) H(0,1;z) \notag \\ &\quad +2 H(0,1;z) H(1-z;y)+4 y H(0,1-z;y)-4 z H(0,1-z;y) \notag \\ &\quad +3 H(0,1-z;y)+2 H(1;z) H(0,z;y)+4 z H(1,0;y)+4 y H(1-z,0;y) \notag \\ &\quad -4 z H(1-z,0;y)+3 H(1-z,0;y)+2 H(1;z) H(1-z,z;y) \notag \\ &\quad -4 y H(z,1-z;y)+2 H(0,z,1-z;y)-2 H(1-z,1,0;y) \notag \\ &\quad +2 H(1-z,z,1-z;y)-2 \zeta_2 H(0;y)-4 y H(1,0;y)-2 H(1,0;y) \notag \\ &\quad -2 H(0,1,0;y)-2 \zeta_2 H(1;z)+4 z H(0,1;z)-4 H(0,1;z)-2 H(0,0,1;z) \notag \\ &\quad +2 H(1,0,1;z)+6 \zeta_3+\zeta_2 \Big ]
+\frac{1}{8\,(y+z)} \Big [ 6 z H(0;y)+2 z H(0;y) H(0;z) \notag \\ &\quad +2 z H(1,0;y)+3 H(0;y) H(1,0;z)+3 \zeta_2 H(0;y)-6 H(0;y)+H(1,0;y) \notag \\ &\quad -3 H(0,1,0;y)+6 \zeta_2 H(1;z)-6 z H(0;z)+2 z H(1,0;z)-H(1,0;z) \notag \\ &\quad +6 H(1,1,0;z)+2 z \zeta_2-6 \zeta_3-\zeta_2 \Big ] -\frac{z}{2\,(1-y)} \Big [ H(1;z) H(z;y) \notag \\ &\quad +H(z,1-z;y) \Big ] -\frac{z^4}{8\,x^2\,(1-y)^2\,(y+z)} \Big [ \zeta_2 H(0;z)+\zeta_2 H(1;z)+H(0,1,0;z) \notag \\ &\quad +H(1,1,0;z)-3 \zeta_3 \Big ] +\frac{z^3}{8\,x^2\,(1-y)\,(y+z)} \Big [ H(1,0;z)+\zeta_2 \Big ] \notag \\ &\quad +\frac{1}{8\,x^2\,(y+z)} \Big [ z^2 \zeta_2 H(0;y)+2 z^2 \zeta_2 H(1;y)-6 z^2 H(0;y) H(0;z) \notag \\ &\quad +z^2 H(1,0;y) H(0;z)-6 z^2 H(1,0;y)+z^2 H(0;y) H(1,0;z)-z^2 H(0,1,0;y) \notag \\ &\quad +2 z^2 H(1,1,0;y)-2 z \zeta_2 H(0;y)+6 z H(0;y) H(0;z)+6 z H(1,0;y) \notag \\ &\quad -2 z H(0;y) H(1,0;z)+2 z H(0,1,0;y)+H(0;y) H(1,0;z)+\zeta_2 H(0;y) \notag \\ &\quad -H(0,1,0;y)+2 z^2 \zeta_2 H(0;z)+3 z^2 \zeta_2 H(1;z)-7 z^2 H(1,0;z) \notag \\ &\quad +3 z^2 H(1,1,0;z)-4 z \zeta_2 H(1;z)+2 \zeta_2 H(1;z)+6 z H(1,0;z)-4 z H(1,1,0;z) \notag \\ &\quad +2 H(1,1,0;z)-7 z^2 \zeta_3-7 z^2 \zeta_2+4 z \zeta_3+6 z \zeta_2-2 \zeta_3 \Big ] \notag \\ &\quad -\frac{3\,z^2}{8\,(1-y)\,y\,(y+z)} H(0;z) H(1-z;y) +\frac{3\,z}{8\,(1-y)\,y} \Big [ H(1;z) H(z;y) \notag \\ &\quad +H(z,1-z;y) \Big ] +\frac{3\,y}{8\,(1-z)\,z} \Big [ -H(0;y) H(1;z)+H(1;z) H(z;y) \notag \\ &\quad -H(0,1-z;y)-H(1-z,0;y)+H(z,1-z;y)+H(1,0;y)+H(0,1;z) \Big ] \\
\mathcal{D}_{2;0,\text{b}}^{(2)} &= \mathcal{D}_{1;0,\text{b}}^{(2)} +\frac{19}{72} -\frac{1}{48} \Big [ H(0;y)+H(0;z) \Big ] -\frac{1}{12\,(1-x)} \Big [ z H(0;y)+y H(0;z) \Big ] \\
\mathcal{D}_{2;0,\text{c}}^{(2)} &= -\frac{5}{32} \zeta_4 + \frac{1}{1296} \Big [ \zeta_2 \Big ( 324 H(0;z) H(1;y) -756 H(1;y)  \notag \\ &\quad +972 H(0;y) H(1;z) +324 H(1;y) H(1;z) +1161 H(1;z) \notag \\ &\quad +648 H(0;z) H(1-z;y) +1296 H(1;z) H(1-z;y) \notag \\ &\quad +1917 H(1-z;y) -1296 H(0,1;y) -324 H(0,1;z) \notag \\ &\quad +972 H(0,1-z;y) -972 H(1,0;y) -324 H(1,0;z) \notag \\ &\quad -1296 H(1,1;y) +324 H(1,1;z) +324 H(1,1-z;y) \notag \\ &\quad +648 H(1-z,0;y) +1296 H(1-z,1-z;y) -2133 \Big ) -540 H(0;y) \notag \\ &\quad -324 H(0;y) H(0;z)+432 H(0;z)-1278 H(0;y) H(1;z)-708 H(1;z) \notag \\ &\quad -1278 H(0;z) H(1-z;y)-1782 H(1;z) H(1-z;y)-708 H(1-z;y) \notag \\ &\quad +4176 H(1;z) H(z;y)+2268 H(1;z) H(0,0;y)+2268 H(1-z;y) H(0,0;z) \notag \\ &\quad -702 H(0;y) H(0,1;z)+1188 H(1-z;y) H(0,1;z)-864 H(z;y) H(0,1;z) \notag \\ &\quad +648 H(0,0;y) H(0,1;z)+2898 H(0,1;z)+594 H(0;z) H(0,1-z;y) \notag \\ &\quad -1188 H(1;z) H(0,1-z;y)+648 H(0,0;z) H(0,1-z;y) \notag \\ &\quad -324 H(0,1;z) H(0,1-z;y)-1278 H(0,1-z;y)-1296 H(1;z) H(0,z;y) \notag \\ &\quad -648 H(0,1;z) H(0,z;y)-540 H(0;z) H(1,0;y)+486 H(1;z) H(1,0;y) \notag \\ &\quad -324 H(0,1;z) H(1,0;y)+1764 H(1,0;y)+1026 H(0;y) H(1,0;z) \notag \\ &\quad +270 H(1-z;y) H(1,0;z)-1296 H(z;y) H(1,0;z) \notag \\ &\quad +648 H(0,0;y) H(1,0;z)-1296 H(0,z;y) H(1,0;z) \notag \\ &\quad -324 H(1,0;y) H(1,0;z)+486 H(1,0;z)-1188 H(0;y) H(1,1;z) \notag \\ &\quad +2376 H(z;y) H(1,1;z)+648 H(0,z;y) H(1,1;z)-1782 H(1,1;z) \notag \\ &\quad -324 H(0,1;z) H(1,1-z;y)+324 H(1,0;z) H(1,1-z;y) \notag \\ &\quad +1566 H(0;z) H(1-z,0;y)-1188 H(1;z) H(1-z,0;y) \notag \\ &\quad +648 H(0,0;z) H(1-z,0;y)+972 H(0,1;z) H(1-z,0;y) \notag \\ &\quad +972 H(1,0;z) H(1-z,0;y)-1278 H(1-z,0;y) \notag \\ &\quad -1188 H(0;z) H(1-z,1-z;y)+1296 H(1,0;z) H(1-z,1-z;y) \notag \\ &\quad -1782 H(1-z,1-z;y)+2376 H(1;z) H(1-z,z;y) \notag \\ &\quad -1944 H(0,1;z) H(1-z,z;y)+648 H(1,0;z) H(1-z,z;y) \notag \\ &\quad -1296 H(1;z) H(z,0;y)+648 H(0,1;z) H(z,0;y)+648 H(1,1;z) H(z,0;y) \notag \\ &\quad -1296 H(0;z) H(z,1-z;y)+2376 H(1;z) H(z,1-z;y) \notag \\ &\quad -648 H(0,1;z) H(z,1-z;y)+648 H(1,0;z) H(z,1-z;y) \notag \\ &\quad +4176 H(z,1-z;y)+432 H(1;z) H(z,z;y)-3240 H(0,1;z) H(z,z;y) \notag \\ &\quad +648 H(1,0;z) H(z,z;y)-1296 H(1,1;z) H(z,z;y) \notag \\ &\quad -648 H(0;y) H(0,0,1;z)-324 H(1;y) H(0,0,1;z) \notag \\ &\quad -1296 H(1-z;y) H(0,0,1;z)-2592 H(z;y) H(0,0,1;z)+108 H(0,0,1;z) \notag \\ &\quad +2268 H(0,0,1-z;y)+648 H(1;z) H(0,0,z;y)-324 H(0;z) H(0,1,0;y) \notag \\ &\quad +324 H(1;z) H(0,1,0;y)-2484 H(0,1,0;y)-324 H(0;y) H(0,1,0;z) \notag \\ &\quad +324 H(1;y) H(0,1,0;z)+1296 H(1-z;y) H(0,1,0;z) \notag \\ &\quad +648 H(z;y) H(0,1,0;z)-1512 H(0,1,0;z)-648 H(z;y) H(0,1,1;z) \notag \\ &\quad +1188 H(0,1,1;z)+2268 H(0,1-z,0;y)-648 H(0;z) H(0,1-z,1-z;y) \notag \\ &\quad -1188 H(0,1-z,1-z;y)+324 H(1;z) H(0,1-z,z;y) \notag \\ &\quad +1296 H(1;z) H(0,z,0;y)-1296 H(0;z) H(0,z,1-z;y) \notag \\ &\quad +648 H(1;z) H(0,z,1-z;y)-1296 H(0,z,1-z;y) \notag \\ &\quad +648 H(1;z) H(0,z,z;y) -648 H(0;z) H(1,0,0;y)+648 H(1;z) H(1,0,0;y) \notag \\ &\quad -2268 H(1,0,0;y)+648 H(1-z;y) H(1,0,0;z)+324 H(0;y) H(1,0,1;z) \notag \\ &\quad -324 H(1;y) H(1,0,1;z)+972 H(1-z;y) H(1,0,1;z) \notag \\ &\quad -648 H(z;y) H(1,0,1;z)+1674 H(1,0,1;z)-648 H(0;z) H(1,0,1-z;y) \notag \\ &\quad +486 H(1,0,1-z;y)+324 H(1;z) H(1,0,z;y)-756 H(1,1,0;y) \notag \\ &\quad +648 H(0;y) H(1,1,0;z)+324 H(1;y) H(1,1,0;z) \notag \\ &\quad +2268 H(1-z;y) H(1,1,0;z)+648 H(z;y) H(1,1,0;z) \notag \\ &\quad -324 H(0;z) H(1,1-z,0;y)+486 H(1,1-z,0;y) \notag \\ &\quad -324 H(1;z) H(1,1-z,z;y)+648 H(0;z) H(1-z,0,0;y) \notag \\ &\quad +2268 H(1-z,0,0;y)+324 H(0;z) H(1-z,0,1-z;y) \notag \\ &\quad -1188 H(1-z,0,1-z;y)+648 H(1;z) H(1-z,0,z;y) \notag \\ &\quad +324 H(0;z) H(1-z,1,0;y)+972 H(1;z) H(1-z,1,0;y) \notag \\ &\quad +1458 H(1-z,1,0;y)+648 H(0;z) H(1-z,1-z,0;y) \notag \\ &\quad -1188 H(1-z,1-z,0;y)+648 H(1;z) H(1-z,z,0;y) \notag \\ &\quad +648 H(0;z) H(1-z,z,1-z;y)+2376 H(1-z,z,1-z;y) \notag \\ &\quad -2592 H(1;z) H(1-z,z,z;y)+648 H(1;z) H(z,0,1-z;y) \notag \\ &\quad -1296 H(z,0,1-z;y)+648 H(1;z) H(z,0,z;y) \notag \\ &\quad +648 H(1;z) H(z,1-z,0;y)-1296 H(z,1-z,0;y) \notag \\ &\quad +648 H(0;z) H(z,1-z,1-z;y)+2376 H(z,1-z,1-z;y) \notag \\ &\quad -1296 H(1;z) H(z,1-z,z;y)+648 H(1;z) H(z,z,0;y) \notag \\ &\quad +648 H(0;z) H(z,z,1-z;y)-1296 H(1;z) H(z,z,1-z;y) \notag \\ &\quad +432 H(z,z,1-z;y)-3888 H(1;z) H(z,z,z;y)-648 H(0,0,1,0;y) \notag \\ &\quad +1296 H(0,0,1,0;z)+648 H(0,0,z,1-z;y)+324 H(0,1,0,1-z;y) \notag \\ &\quad -1296 H(0,1,1,0;y)+648 H(0,1,1,0;z)+324 H(0,1,1-z,0;y) \notag \\ &\quad +972 H(0,1-z,1,0;y)+324 H(0,1-z,z,1-z;y) \notag \\ &\quad +1296 H(0,z,0,1-z;y)+1296 H(0,z,1-z,0;y)+648 H(0,z,1-z,1-z;y) \notag \\ &\quad +648 H(0,z,z,1-z;y)-324 H(1,0,0,1;z)+648 H(1,0,0,1-z;y) \notag \\ &\quad -1296 H(1,0,1,0;y)+648 H(1,0,1,0;z)+648 H(1,0,1-z,0;y) \notag \\ &\quad +324 H(1,0,z,1-z;y)-1296 H(1,1,0,0;y)+648 H(1,1,0,1;z) \notag \\ &\quad -1296 H(1,1,1,0;y)+972 H(1,1,1,0;z)+648 H(1,1-z,0,0;y) \notag \\ &\quad -324 H(1,1-z,z,1-z;y)+648 H(1-z,0,1,0;y) \notag \\ &\quad +648 H(1-z,0,z,1-z;y)+648 H(1-z,1,0,0;y) \notag \\ &\quad +972 H(1-z,1,0,1-z;y)+972 H(1-z,1,1-z,0;y) \notag \\ &\quad +1296 H(1-z,1-z,1,0;y)+648 H(1-z,z,0,1-z;y) \notag \\ &\quad +648 H(1-z,z,1-z,0;y)-2592 H(1-z,z,z,1-z;y) \notag \\ &\quad +648 H(z,0,1-z,1-z;y)+648 H(z,0,z,1-z;y) \notag \\ &\quad +648 H(z,1-z,0,1-z;y)+648 H(z,1-z,1-z,0;y) \notag \\ &\quad -1296 H(z,1-z,z,1-z;y)+648 H(z,z,0,1-z;y)+648 H(z,z,1-z,0;y) \notag \\ &\quad -1296 H(z,z,1-z,1-z;y)-3888 H(z,z,z,1-z;y)-972 \zeta_3 H(1;y) \notag \\ &\quad +648  \zeta_3 H(1;z) +1620 \zeta_3 H(1-z;y) +1845 \zeta_3-38 \Big ] \notag \\ &\quad + \frac{3\,z}{4\,x} \Big [ -H(0;y) H(0;z)-H(0;y)-H(1,0;y)+H(0;z) \notag \\ &\quad -H(1,0;z)-\zeta_2 \Big ] - \frac{3}{4\,(1-z)} H(0;z) + \frac{3\,z}{4\,x^2} \Big [ -z H(0;y) H(0;z) \notag \\ &\quad +H(0;y) H(0;z)-z H(1,0;y)+H(1,0;y)-z H(1,0;z)+H(1,0;z) \notag \\ &\quad -z \zeta_2+\zeta_2 \Big ] + \frac{3\,z}{8\,y} \Big [ H(0;z) H(1-z;y)-H(1;z) H(z;y)-H(z,1-z;y) \Big ] \notag \\ &\quad + \frac{3}{8\,z} \Big [ y H(0;y) H(1;z)-y H(1;z) H(z;y)-y H(0,1;z)+y H(0,1-z;y) \notag \\ &\quad +y H(1-z,0;y)-y H(z,1-z;y)-y H(1,0;y) \Big ] + \frac{3 z}{4\,x\,(1-y)} H(0;y) \\
\mathcal{D}_{2;0,\text{d}}^{(2)} &= \mathcal{D}_{1;0,\text{d}}^{(2)} \\
\mathcal{D}_{2;0,\text{e}}^{(2)} &= \mathcal{D}_{1;0,\text{e}}^{(2)} + \frac{(1-z)\,z}{12\,(1-x)\,x^2} \Big [ H(0;y) H(0;z)+H(1,0;y) +H(1,0;z)+\zeta_2 \Big ] \notag \\ &\quad - \frac{1}{12\,(1-x)\,x} \Big [ z H(0;y)+z H(0;y) H(0;z)+z H(1,0;y) -H(0;y) \notag \\ &\quad -z H(0;z)+z H(1,0;z)+z \zeta_2 \Big ] -\frac{1}{48\,(1-x)} \Big [ 5 H(0;y) -x H(0;y) \notag \\ &\quad +4 z H(0;y) -4 z H(0;z) \Big ] + \frac{1}{144} \Big [ 74 -15 H(0;z) \Big ] \\
\mathcal{D}_{2;0,\text{f}}^{(2)} &= \mathcal{D}_{1;0,\text{f}}^{(2)} +\frac{1}{288} \Big [ 36 H(0;y) H(0;z)+60 H(0;y)+36 H(1,0;y)+192 H(0;z) \notag \\ &\quad +36 H(1,0;z)+36 \zeta_2-565 \Big ] +\frac{3}{4\,(1-z)\,(y+z)} H(0;z) \notag \\ &\quad +\frac{1}{24\,x\,(y+z)} \Big [ 12 z^2 H(0;y) H(0;z)+12 z^2 H(1,0;y)+6 z \zeta_2 H(0;y) \notag \\ &\quad -24 z \zeta_2 H(1;y)+20 z H(0;y)+8 z H(0;y) H(0;z)-12 z H(1,0;y) H(0;z) \notag \\ &\quad +2 z H(1,0;y)+6 z H(0;y) H(1,0;z)-6 z H(0,1,0;y)-24 z H(1,1,0;y) \notag \\ &\quad -6 H(0;y) H(1,0;z)-6 \zeta_2 H(0;y)-2 H(0;y)+3 H(1,0;y)+6 H(0,1,0;y) \notag \\ &\quad +12 z^2 H(1,0;z)-12 z \zeta_2 H(0;z)+12 z \zeta_2 H(1;z)-12 \zeta_2 H(1;z)-20 z H(0;z) \notag \\ &\quad +14 z H(1,0;z)+12 z H(0,1,0;z)+12 z H(1,1,0;z)-3 H(1,0;z) \notag \\ &\quad -12 H(1,1,0;z)+12 z^2 \zeta_2+12 z \zeta_3+8 z \zeta_2+12 \zeta_3-3 \zeta_2 \Big ] \notag \\ &\quad +\frac{1}{24\,(y+z)} \Big [ 11 z H(0;y)+12 z H(0;y) H(0;z)+12 z H(1,0;y) \notag \\ &\quad +9 H(0;y) H(1,0;z)+9 \zeta_2 H(0;y)+2 H(0;y)-3 H(1,0;y)-9 H(0,1,0;y) \notag \\ &\quad +18 \zeta_2 H(1;z)-11 z H(0;z)-18 H(0;z)+12 z H(1,0;z)+3 H(1,0;z) \notag \\ &\quad +18 H(1,1,0;z)+12 z \zeta_2-18 \zeta_3+3 \zeta_2 \Big ] +\frac{3\,z^2}{4\,x^2\,(1-y)\,(y+z)} H(0;y) \notag \\ &\quad +\frac{1}{24\,x^2\,(y+z)} \Big [ -3 z^2 \zeta_2 H(0;y)-6 z^2 \zeta_2 H(1;y)+20 z^2 H(0;y) H(0;z) \notag \\ &\quad -3 z^2 H(1,0;y) H(0;z)+20 z^2 H(1,0;y)-3 z^2 H(0;y) H(1,0;z) \notag \\ &\quad +3 z^2 H(0,1,0;y)-6 z^2 H(1,1,0;y)+6 z \zeta_2 H(0;y)-18 z H(0;y) \notag \\ &\quad -20 z H(0;y) H(0;z)-20 z H(1,0;y)+6 z H(0;y) H(1,0;z)-6 z H(0,1,0;y) \notag \\ &\quad -3 H(0;y) H(1,0;z)-3 \zeta_2 H(0;y)+3 H(0,1,0;y)-3 z^2 \zeta_2 H(0;z) \notag \\ &\quad -6 z^2 \zeta_2 H(1;z)+20 z^2 H(1,0;z)+3 z^2 H(0,1,0;z)-6 z^2 H(1,1,0;z) \notag \\ &\quad +12 z \zeta_2 H(1;z)-6 \zeta_2 H(1;z)-20 z H(1,0;z)+12 z H(1,1,0;z) \notag \\ &\quad -6 H(1,1,0;z)+12 z^2 \zeta_3+20 z^2 \zeta_2-12 z \zeta_3-20 z \zeta_2+6 \zeta_3 \Big ] \, .
\end{align} 

\section{Formulae for soft and collinear limits}
\label{matrixelements}

We list the unrenormalized $H \to b \overline{b}$ matrix elements that are needed for the soft and collinear limit checks of the two-loop \hbbg amplitude. The matrix elements in CDR read:
\begin{align}
\mathcal{M}^{(0)}_{H \to b \bar{b}} \mathcal{M}^{(0)*}_{H \to b \bar{b}} &= 2 \, y_b^2\, m_H^2\, N_c \\
\mathcal{M}^{(1)}_{H \to b \bar{b}} \mathcal{M}^{(0)*}_{H \to b \bar{b}} &= -\left ( \frac{\as}{2\pi} \right )\, \mathcal{M}^{(0)}_{H \to b \bar{b}} \mathcal{M}^{(0)*}_{H \to b \bar{b}} \, (4\pi)^{\eps} S_{\epsilon} \,C_F \notag \\ &\quad \left (-\frac{m_H^2}{\mu^2}\right )^{-\epsilon} \, \frac{(1-\epsilon)^2}{\epsilon^2} \frac{\Gamma(1+\epsilon) \,\Gamma(1-\epsilon)^2}{\Gamma(2-2\epsilon)} \\
\mathcal{M}^{(2)}_{H \to b \bar{b}} \mathcal{M}^{(0)*}_{H \to b \bar{b}} &= \frac{1}{4} \,\left ( \frac{\as}{2\pi} \right )^2\, \mathcal{M}^{(0)}_{H \to b \bar{b}} \mathcal{M}^{(0)*}_{H \to b \bar{b}} \, \left (-\frac{m_H^2}{\mu^2}\right )^{-2\epsilon} \, \mathcal{F}_2
\end{align}
where $\mathcal{F}_2$ is taken from Eq.~(2.24) of Ref.~\cite{Gehrmann:2014vha} and $S_{\epsilon} = \frac{\exp{(\epsilon \gamma_E)}}{(4\pi)^{\epsilon}}$.

\vspace{5mm}

The soft currents in Eq.~\eqref{softlimiteq} are defined as
\begin{align}
S^{(0)}(y,z) &= 16\pi \as \, C_F\, \frac{1}{m_H^2} \left ( \frac{1-y-z}{y \,z} \right ) \\
S^{(1)}(y,z) &= -\frac{1}{2} S^{(0)}(y,z) \left ( \frac{\as}{2\pi} \right ) (4\pi)^{\eps} S_{\epsilon} \,C_A \left (-\frac{m_H^2}{\mu^2}\right )^{-\epsilon} \notag \\ &\quad \left ( \frac{1-y-z}{y\, z} \right )^{\epsilon} \frac{\Gamma(1-\epsilon)^3 \,\Gamma(1+\epsilon)^2}{\epsilon^2\, \Gamma(1-2\epsilon)} \\
S^{(2)}(y,z) &= \frac{1}{4} S^{(0)}(y,z) \left ( \frac{\as}{2\pi} \right )^2 \, \left (-\frac{m_H^2}{\mu^2}\right )^{-2\epsilon} \left ( \frac{1-y-z}{y\, z} \right )^{2\epsilon} \notag \\ &\quad \bigg [ C_A N_f \left ( \frac{1}{6 \eps^3}+\frac{5}{18 \eps^2} + \frac{1}{\eps} \left (\frac{19}{54} + \frac{\zeta_2}{6} \right ) + \frac{65}{162} + \frac{5 \zeta_2}{18} -\frac{31 \zeta_3}{9} \right ) \notag \\ &\quad + C_A^2 \bigg ( \frac{1}{2 \eps^4}-\frac{11}{12 \eps^3} +\frac{1}{\eps^2} \left (-\frac{67}{36}+\zeta_2 \right ) +\frac{1}{\eps} \left (-\frac{193}{54}-\frac{11 \zeta_2}{12}-\frac{11 \zeta_3}{6} \right ) \notag \\ &\quad -\frac{571}{81} -\frac{67 \zeta_2}{36} + \frac{341 \zeta_3}{18} +\frac{7 \zeta_4}{8} \bigg ) + \mathcal{O}(\eps) \bigg ]
\end{align}
where $S^{(0)}(y,z)$ and $S^{(1)}(y,z)$ have been adapted from Eqs.~(12), (13), and (26) of Ref.~\cite{Catani:2000pi}, while $S^{(2)}(y,z)$ is taken from Eq.~(11) of Ref.~\cite{Li:2013lsa}.

\vspace{5mm}

The collinear functions in Eq.~\eqref{collinearlimiteq} are
\begin{align}
C^{(0)}(y,z) &= 4\pi \as \, C_F\, \sum_{n=1}^2 \text{Sp}^{(0)}_n(\epsilon) \\
C^{(1)}(y,z) &= \frac{1}{2} \frac{1}{(2\pi)^2} (4\pi \as)^2 \, S_{\epsilon}\, C_F\, \sum_{n=1}^2 \text{Sp}^{(1)}_n(\epsilon) \, \text{Sp}^{(0)}_n(\epsilon) \\
C^{(2)}(y,z) &= \frac{1}{2} \frac{1}{(2\pi)^4} (4\pi \as)^3 \, S_{\epsilon}^2\, C_F \, \sum_{n=1}^2 \text{Sp}^{(2)}_n(\epsilon) \, \text{Sp}^{(0)}_n(\epsilon) \, .
\end{align}
The tree-level splitting functions in CDR are (see Eqs.~(4.17) and (4.18) of Ref.~\cite{Badger:2004uk}):
\begin{align}
\text{Sp}^{(0)}_1(\epsilon) &= \frac{1}{m_H^2} \frac{2}{y} \left [ \frac{1+\epsilon (1-2 z)}{z} \right ] \\
\text{Sp}^{(0)}_2(\epsilon) &= \frac{1}{m_H^2} \frac{2}{y} \left [ (1-\epsilon) \frac{(1-z)^2}{z} \right ] \, .
\end{align}
At one loop, the splitting functions in CDR are (see Eqs.~(4.21) and (4.22) of Ref.~\cite{Badger:2004uk}):
\begin{align}
\text{Sp}^{(1)}_1(\epsilon) &= -c_{\Gamma}(\epsilon) \, (4\pi)^{\eps} \, \left (-\frac{m_H^2}{\mu^2}\right )^{-\epsilon} y^{-\epsilon} \bigg [ s^{(1)}(\eps) -\left ( \frac{N_c^2+1}{N_c} \right ) \frac{z}{4 (1-2\epsilon)} \bigg ] \\
\text{Sp}^{(1)}_2(\epsilon) &= -c_{\Gamma}(\epsilon) \, (4\pi)^{\eps} \, \left (-\frac{m_H^2}{\mu^2}\right )^{-\epsilon} y^{-\epsilon} \, s^{(1)}(\eps) \, ,
\end{align}
with
\begin{align}
c_{\Gamma}(\epsilon) &= \frac{\Gamma(1-\epsilon)^2 \, \Gamma(1+\epsilon)}{\Gamma(1-2\epsilon)} \\
s^{(1)}(\eps) &=  \frac{N_c}{2 \epsilon^2} \bigg [ 1-\sum_{m=1}^{\infty} \epsilon^m \bigg ( \text{Li}_m \left (\frac{1-z}{-z} \right ) -\frac{1}{N_c^2}\text{Li}_m \left (\frac{-z}{1-z}\right ) \bigg ) \bigg ] \, .
\end{align}
The two-loop splitting functions $\text{Sp}^{(2)}_n(\epsilon)$ in CDR are (see Eqs.~(3.7)$-$(3.14) and (4.24)$-$(4.25) of Ref.~\cite{Badger:2004uk}):
\begin{align}
\text{Sp}^{(2)}_n(\epsilon) &= \frac{1}{2} \left ( \text{Sp}^{(1)}_n(\epsilon) \right )^2 + \frac{e^{-\epsilon \gamma_E}\,c_{\Gamma}(\epsilon)}{c_{\Gamma}(2\epsilon)} \left (\frac{\beta_0}{2 \epsilon} + K \right ) \text{Sp}^{(1)}_n(2\epsilon) + H_{q\bar{q}g}(\epsilon) + \text{Sp}^{(2),\text{fin}}_n + \mathcal{O}(\epsilon) \, ,
\end{align}
where
\begin{align}
H_{q\bar{q}g}(\epsilon) = \frac{e^{-\epsilon \gamma_E}\,c_{\Gamma}(\epsilon)}{\epsilon} (4\pi)^{2\eps} \, \left (-\frac{m_H^2}{\mu^2}\right )^{-2\epsilon} y^{-2\epsilon} \left [ z (1-z) \right ]^{-2\epsilon} \left ( H^{(2)}_g-\frac{\beta_0}{8} K + \frac{\beta_1}{16} \right )
\end{align}
with $H^{(2)}_g$ and $K$ as defined in Appendix \ref{renoirformulae}. Finally, the functions $\text{Sp}^{(2),\text{fin}}_1$ and $\text{Sp}^{(2),\text{fin}}_2$ correspond to Eqs.~(4.24) and (4.25) of Ref.~\cite{Badger:2004uk} respectively with the replacement $w \to z$.

\bibliographystyle{JHEP}
\bibliography{HbbgNNLO_MW}

\providecommand{\href}[2]{#2}\begingroup\raggedright\begin{thebibliography}{10}

\bibitem{Aad:2012tfa}
{\bf ATLAS} Collaboration, G.~Aad {\em et.~al.}, {\it {Observation of a new
  particle in the search for the Standard Model Higgs boson with the ATLAS
  detector at the LHC}},  {\em Phys. Lett.} {\bf B716} (2012) 1--29
  [\href{http://arXiv.org/abs/1207.7214}{{\tt 1207.7214}}].

\bibitem{Chatrchyan:2012xdj}
{\bf CMS} Collaboration, S.~Chatrchyan {\em et.~al.}, {\it {Observation of a
  new boson at a mass of 125 GeV with the CMS experiment at the LHC}},  {\em
  Phys. Lett.} {\bf B716} (2012) 30--61
  [\href{http://arXiv.org/abs/1207.7235}{{\tt 1207.7235}}].

\bibitem{Baer:2013cma}
H.~Baer, T.~Barklow, K.~Fujii, Y.~Gao, A.~Hoang, S.~Kanemura, J.~List, H.~E.
  Logan, A.~Nomerotski, M.~Perelstein {\em et.~al.}, {\it {The International
  Linear Collider Technical Design Report - Volume 2: Physics}},
  \href{http://arXiv.org/abs/1306.6352}{{\tt 1306.6352}}.

\bibitem{Gomez-Ceballos:2013zzn}
{\bf TLEP Design Study Working Group} Collaboration, M.~Bicer {\em et.~al.},
  {\it {First Look at the Physics Case of TLEP}},  {\em JHEP} {\bf 01} (2014)
  164 [\href{http://arXiv.org/abs/1308.6176}{{\tt 1308.6176}}].

\bibitem{Abada:2019zxq}
{\bf FCC} Collaboration, A.~Abada {\em et.~al.}, {\it {FCC-ee: The Lepton
  Collider}}, . [Eur. Phys. J. ST228,no.2,261(2019)].

\bibitem{Aaboud:2018zhk}
{\bf ATLAS} Collaboration, M.~Aaboud {\em et.~al.}, {\it {Observation of $H
  \rightarrow b\bar{b}$ decays and $VH$ production with the ATLAS detector}},
  {\em Phys. Lett.} {\bf B786} (2018) 59--86
  [\href{http://arXiv.org/abs/1808.08238}{{\tt 1808.08238}}].

\bibitem{Sirunyan:2018kst}
{\bf CMS} Collaboration, A.~M. Sirunyan {\em et.~al.}, {\it {Observation of
  Higgs boson decay to bottom quarks}},  {\em Phys. Rev. Lett.} {\bf 121}
  (2018), no.~12 121801 [\href{http://arXiv.org/abs/1808.08242}{{\tt
  1808.08242}}].

\bibitem{Sirunyan:2017dgcx}
{\bf CMS} Collaboration, A.~M. Sirunyan {\em et.~al.}, {\it {Inclusive search
  for a highly boosted Higgs boson decaying to a bottom quark-antiquark pair}},
   {\em Phys. Rev. Lett.} {\bf 120} (2018), no.~7 071802
  [\href{http://arXiv.org/abs/1709.05543}{{\tt 1709.05543}}].

\bibitem{Anastasiou:2015ema}
C.~Anastasiou, C.~Duhr, F.~Dulat, F.~Herzog and B.~Mistlberger, {\it {Higgs
  Boson Gluon-Fusion Production in QCD at Three Loops}},  {\em Phys. Rev.
  Lett.} {\bf 114} (2015) 212001 [\href{http://arXiv.org/abs/1503.06056}{{\tt
  1503.06056}}].

\bibitem{Anastasiou:2016cez}
C.~Anastasiou, C.~Duhr, F.~Dulat, E.~Furlan, T.~Gehrmann, F.~Herzog,
  A.~Lazopoulos and B.~Mistlberger, {\it {High precision determination of the
  gluon fusion Higgs boson cross-section at the LHC}},  {\em JHEP} {\bf 05}
  (2016) 058 [\href{http://arXiv.org/abs/1602.00695}{{\tt 1602.00695}}].

\bibitem{Catani:2007vq}
S.~Catani and M.~Grazzini, {\it {An NNLO subtraction formalism in hadron
  collisions and its application to Higgs boson production at the LHC}},  {\em
  Phys. Rev. Lett.} {\bf 98} (2007) 222002
  [\href{http://arXiv.org/abs/hep-ph/0703012}{{\tt hep-ph/0703012}}].

\bibitem{Cieri:2018oms}
L.~Cieri, X.~Chen, T.~Gehrmann, E.~W.~N. Glover and A.~Huss, {\it {Higgs boson
  production at the LHC using the $q_T$ subtraction formalism at N$^3$LO QCD}},
   {\em JHEP} {\bf 02} (2019) 096 [\href{http://arXiv.org/abs/1807.11501}{{\tt
  1807.11501}}].

\bibitem{Dulat:2017prg}
F.~Dulat, B.~Mistlberger and A.~Pelloni, {\it {Differential Higgs production at
  N$^{3}$LO beyond threshold}},  {\em JHEP} {\bf 01} (2018) 145
  [\href{http://arXiv.org/abs/1710.03016}{{\tt 1710.03016}}].

\bibitem{Dulat:2018bfe}
F.~Dulat, B.~Mistlberger and A.~Pelloni, {\it {Precision predictions at N$^3$LO
  for the Higgs boson rapidity distribution at the LHC}},  {\em Phys. Rev.}
  {\bf D99} (2019), no.~3 034004 [\href{http://arXiv.org/abs/1810.09462}{{\tt
  1810.09462}}].

\bibitem{Boughezal:2013uia}
R.~Boughezal, F.~Caola, K.~Melnikov, F.~Petriello and M.~Schulze, {\it {Higgs
  boson production in association with a jet at next-to-next-to-leading order
  in perturbative QCD}},  {\em JHEP} {\bf 06} (2013) 072
  [\href{http://arXiv.org/abs/1302.6216}{{\tt 1302.6216}}].

\bibitem{Chen:2014gva}
X.~Chen, T.~Gehrmann, E.~W.~N. Glover and M.~Jaquier, {\it {Precise QCD
  predictions for the production of Higgs + jet final states}},  {\em Phys.
  Lett.} {\bf B740} (2015) 147--150 [\href{http://arXiv.org/abs/1408.5325}{{\tt
  1408.5325}}].

\bibitem{Campbell:2006xx}
J.~M. Campbell, R.~K. Ellis and G.~Zanderighi, {\it {Next-to-Leading order
  Higgs + 2 jet production via gluon fusion}},  {\em JHEP} {\bf 10} (2006) 028
  [\href{http://arXiv.org/abs/hep-ph/0608194}{{\tt hep-ph/0608194}}].

\bibitem{Cullen:2013saa}
G.~Cullen, H.~van Deurzen, N.~Greiner, G.~Luisoni, P.~Mastrolia, E.~Mirabella,
  G.~Ossola, T.~Peraro and F.~Tramontano, {\it {Next-to-Leading-Order QCD
  Corrections to Higgs Boson Production Plus Three Jets in Gluon Fusion}},
  {\em Phys. Rev. Lett.} {\bf 111} (2013), no.~13 131801
  [\href{http://arXiv.org/abs/1307.4737}{{\tt 1307.4737}}].

\bibitem{Gehrmann:2011aa}
T.~Gehrmann, M.~Jaquier, E.~W.~N. Glover and A.~Koukoutsakis, {\it {Two-Loop
  QCD Corrections to the Helicity Amplitudes for $H \to$ 3 partons}},  {\em
  JHEP} {\bf 02} (2012) 056 [\href{http://arXiv.org/abs/1112.3554}{{\tt
  1112.3554}}].

\bibitem{Braaten:1980yq}
E.~Braaten and J.~P. Leveille, {\it {Higgs Boson Decay and the Running Mass}},
  {\em Phys. Rev.} {\bf D22} (1980) 715.

\bibitem{Anastasiou:2011qx}
C.~Anastasiou, F.~Herzog and A.~Lazopoulos, {\it {The fully differential decay
  rate of a Higgs boson to bottom-quarks at NNLO in QCD}},  {\em JHEP} {\bf 03}
  (2012) 035 [\href{http://arXiv.org/abs/1110.2368}{{\tt 1110.2368}}].

\bibitem{DelDuca:2015zqa}
V.~Del~Duca, C.~Duhr, G.~Somogyi, F.~Tramontano and Z.~Tr\'ocs\'anyi, {\it
  {Higgs boson decay into b-quarks at NNLO accuracy}},  {\em JHEP} {\bf 04}
  (2015) 036 [\href{http://arXiv.org/abs/1501.07226}{{\tt 1501.07226}}].

\bibitem{Bernreuther:2018ynm}
W.~Bernreuther, L.~Chen and Z.-G. Si, {\it {Differential decay rates of CP-even
  and CP-odd Higgs bosons to top and bottom quarks at NNLO QCD}},  {\em JHEP}
  {\bf 07} (2018) 159 [\href{http://arXiv.org/abs/1805.06658}{{\tt
  1805.06658}}].

\bibitem{Baikov:2005rw}
P.~A. Baikov, K.~G. Chetyrkin and J.~H. Kuhn, {\it {Scalar correlator at
  O(alpha(s)**4), Higgs decay into b-quarks and bounds on the light quark
  masses}},  {\em Phys. Rev. Lett.} {\bf 96} (2006) 012003
  [\href{http://arXiv.org/abs/hep-ph/0511063}{{\tt hep-ph/0511063}}].

\bibitem{Ahmed:2014pka}
T.~Ahmed, M.~Mahakhud, P.~Mathews, N.~Rana and V.~Ravindran, {\it {Two-loop QCD
  corrections to Higgs $\to b+\overline{b}+g$ amplitude}},  {\em JHEP} {\bf 08}
  (2014) 075 [\href{http://arXiv.org/abs/1405.2324}{{\tt 1405.2324}}].

\bibitem{Mondini:2019gid}
R.~Mondini, M.~Schiavi and C.~Williams, {\it {N$^{3}$LO predictions for the
  decay of the Higgs boson to bottom quarks}},
  \href{http://arXiv.org/abs/1904.08960}{{\tt 1904.08960}}.

\bibitem{Chetyrkin:1996sr}
K.~G. Chetyrkin, {\it {Correlator of the quark scalar currents and Gamma(tot)
  (H ---> hadrons) at O (alpha-s**3) in pQCD}},  {\em Phys. Lett.} {\bf B390}
  (1997) 309--317 [\href{http://arXiv.org/abs/hep-ph/9608318}{{\tt
  hep-ph/9608318}}].

\bibitem{Gehrmann:2014vha}
T.~Gehrmann and D.~Kara, {\it {The $Hb\bar{b}$ form factor to three loops in
  QCD}},  {\em JHEP} {\bf 09} (2014) 174
  [\href{http://arXiv.org/abs/1407.8114}{{\tt 1407.8114}}].

\bibitem{Primo:2018zby}
A.~Primo, G.~Sasso, G.~Somogyi and F.~Tramontano, {\it {Exact Top Yukawa
  corrections to Higgs boson decay into bottom quarks}},
  \href{http://arXiv.org/abs/1812.07811}{{\tt 1812.07811}}.

\bibitem{Gaunt:2015pea}
J.~Gaunt, M.~Stahlhofen, F.~J. Tackmann and J.~R. Walsh, {\it {N-jettiness
  Subtractions for NNLO QCD Calculations}},  {\em JHEP} {\bf 09} (2015) 058
  [\href{http://arXiv.org/abs/1505.04794}{{\tt 1505.04794}}].

\bibitem{Boughezal:2015dva}
R.~Boughezal, C.~Focke, X.~Liu and F.~Petriello, {\it {$W$-boson production in
  association with a jet at next-to-next-to-leading order in perturbative
  QCD}},  {\em Phys. Rev. Lett.} {\bf 115} (2015), no.~6 062002
  [\href{http://arXiv.org/abs/1504.02131}{{\tt 1504.02131}}].

\bibitem{Stewart:2010tn}
I.~W. Stewart, F.~J. Tackmann and W.~J. Waalewijn, {\it {N-Jettiness: An
  Inclusive Event Shape to Veto Jets}},  {\em Phys. Rev. Lett.} {\bf 105}
  (2010) 092002 [\href{http://arXiv.org/abs/1004.2489}{{\tt 1004.2489}}].

\bibitem{Brown:1990nm}
N.~Brown and W.~J. Stirling, {\it {Jet cross-sections at leading double
  logarithm in e+ e- annihilation}},  {\em Phys. Lett.} {\bf B252} (1990)
  657--662.

\bibitem{Catani:1991hj}
S.~Catani, Y.~L. Dokshitzer, M.~Olsson, G.~Turnock and B.~R. Webber, {\it {New
  clustering algorithm for multi - jet cross-sections in e+ e- annihilation}},
  {\em Phys. Lett.} {\bf B269} (1991) 432--438.

\bibitem{Stewart:2009yx}
I.~W. Stewart, F.~J. Tackmann and W.~J. Waalewijn, {\it {Factorization at the
  LHC: From PDFs to Initial State Jets}},  {\em Phys. Rev.} {\bf D81} (2010)
  094035 [\href{http://arXiv.org/abs/0910.0467}{{\tt 0910.0467}}].

\bibitem{Becher:2006qw}
T.~Becher and M.~Neubert, {\it {Toward a NNLO calculation of the anti-B --->
  X(s) gamma decay rate with a cut on photon energy. II. Two-loop result for
  the jet function}},  {\em Phys. Lett.} {\bf B637} (2006) 251--259
  [\href{http://arXiv.org/abs/hep-ph/0603140}{{\tt hep-ph/0603140}}].

\bibitem{Campbell:2017hsw}
J.~M. Campbell, R.~K. Ellis, R.~Mondini and C.~Williams, {\it {The NNLO QCD
  soft function for 1-jettiness}},  {\em Eur. Phys. J.} {\bf C78} (2018), no.~3
  234 [\href{http://arXiv.org/abs/1711.09984}{{\tt 1711.09984}}].

\bibitem{Boughezal:2015eha}
R.~Boughezal, X.~Liu and F.~Petriello, {\it {$N$-jettiness soft function at
  next-to-next-to-leading order}},  {\em Phys. Rev.} {\bf D91} (2015), no.~9
  094035 [\href{http://arXiv.org/abs/1504.02540}{{\tt 1504.02540}}].

\bibitem{Bern:1994zx}
Z.~Bern, L.~J. Dixon, D.~C. Dunbar and D.~A. Kosower, {\it {One loop n point
  gauge theory amplitudes, unitarity and collinear limits}},  {\em Nucl. Phys.}
  {\bf B425} (1994) 217--260 [\href{http://arXiv.org/abs/hep-ph/9403226}{{\tt
  hep-ph/9403226}}].

\bibitem{Britto:2004nc}
R.~Britto, F.~Cachazo and B.~Feng, {\it {Generalized unitarity and one-loop
  amplitudes in N=4 super-Yang-Mills}},  {\em Nucl. Phys.} {\bf B725} (2005)
  275--305 [\href{http://arXiv.org/abs/hep-th/0412103}{{\tt hep-th/0412103}}].

\bibitem{Forde:2007mi}
D.~Forde, {\it {Direct extraction of one-loop integral coefficients}},  {\em
  Phys. Rev.} {\bf D75} (2007) 125019
  [\href{http://arXiv.org/abs/0704.1835}{{\tt 0704.1835}}].

\bibitem{Mastrolia:2009dr}
P.~Mastrolia, {\it {Double-Cut of Scattering Amplitudes and Stokes' Theorem}},
  {\em Phys. Lett.} {\bf B678} (2009) 246--249
  [\href{http://arXiv.org/abs/0905.2909}{{\tt 0905.2909}}].

\bibitem{Badger:2008cm}
S.~D. Badger, {\it {Direct Extraction Of One Loop Rational Terms}},  {\em JHEP}
  {\bf 01} (2009) 049 [\href{http://arXiv.org/abs/0806.4600}{{\tt 0806.4600}}].

\bibitem{Ellis:2008ir}
R.~K. Ellis, W.~T. Giele, Z.~Kunszt and K.~Melnikov, {\it {Masses, fermions and
  generalized $D$-dimensional unitarity}},  {\em Nucl. Phys.} {\bf B822} (2009)
  270--282 [\href{http://arXiv.org/abs/0806.3467}{{\tt 0806.3467}}].

\bibitem{Badger:2009hw}
S.~Badger, E.~W. Nigel~Glover, P.~Mastrolia and C.~Williams, {\it {One-loop
  Higgs plus four gluon amplitudes: Full analytic results}},  {\em JHEP} {\bf
  01} (2010) 036 [\href{http://arXiv.org/abs/0909.4475}{{\tt 0909.4475}}].

\bibitem{Britto:2005fq}
R.~Britto, F.~Cachazo, B.~Feng and E.~Witten, {\it {Direct proof of tree-level
  recursion relation in Yang-Mills theory}},  {\em Phys. Rev. Lett.} {\bf 94}
  (2005) 181602 [\href{http://arXiv.org/abs/hep-th/0501052}{{\tt
  hep-th/0501052}}].

\bibitem{Alwall:2011uj}
J.~Alwall, M.~Herquet, F.~Maltoni, O.~Mattelaer and T.~Stelzer, {\it {MadGraph
  5 : Going Beyond}},  {\em JHEP} {\bf 06} (2011) 128
  [\href{http://arXiv.org/abs/1106.0522}{{\tt 1106.0522}}].

\bibitem{Catani:1996vz}
S.~Catani and M.~H. Seymour, {\it {A General algorithm for calculating jet
  cross-sections in NLO QCD}},  {\em Nucl. Phys.} {\bf B485} (1997) 291--419
  [\href{http://arXiv.org/abs/hep-ph/9605323}{{\tt hep-ph/9605323}}]. [Erratum:
  Nucl. Phys.B510,503(1998)].

\bibitem{Becher:2013vva}
T.~Becher, G.~Bell, C.~Lorentzen and S.~Marti, {\it {Transverse-momentum
  spectra of electroweak bosons near threshold at NNLO}},  {\em JHEP} {\bf 02}
  (2014) 004 [\href{http://arXiv.org/abs/1309.3245}{{\tt 1309.3245}}].

\bibitem{Hahn:2000kx}
T.~Hahn, {\it {Generating Feynman diagrams and amplitudes with FeynArts 3}},
  {\em Comput. Phys. Commun.} {\bf 140} (2001) 418--431
  [\href{http://arXiv.org/abs/hep-ph/0012260}{{\tt hep-ph/0012260}}].

\bibitem{Maierhoefer:2017hyi}
P.~Maierhöfer, J.~Usovitsch and P.~Uwer, {\it {Kira—A Feynman integral
  reduction program}},  {\em Comput. Phys. Commun.} {\bf 230} (2018) 99--112
  [\href{http://arXiv.org/abs/1705.05610}{{\tt 1705.05610}}].

\bibitem{Lee:2013mka}
R.~N. Lee, {\it {LiteRed 1.4: a powerful tool for reduction of multiloop
  integrals}},  {\em J. Phys. Conf. Ser.} {\bf 523} (2014) 012059
  [\href{http://arXiv.org/abs/1310.1145}{{\tt 1310.1145}}].

\bibitem{Garland:2001tf}
L.~W. Garland, T.~Gehrmann, E.~W.~N. Glover, A.~Koukoutsakis and E.~Remiddi,
  {\it {The Two loop QCD matrix element for e+ e- ---> 3 jets}},  {\em Nucl.
  Phys.} {\bf B627} (2002) 107--188
  [\href{http://arXiv.org/abs/hep-ph/0112081}{{\tt hep-ph/0112081}}].

\bibitem{Remiddi:1999ew}
E.~Remiddi and J.~A.~M. Vermaseren, {\it {Harmonic polylogarithms}},  {\em Int.
  J. Mod. Phys.} {\bf A15} (2000) 725--754
  [\href{http://arXiv.org/abs/hep-ph/9905237}{{\tt hep-ph/9905237}}].

\bibitem{Gehrmann:2000zt}
T.~Gehrmann and E.~Remiddi, {\it {Two loop master integrals for gamma* ---> 3
  jets: The Planar topologies}},  {\em Nucl. Phys.} {\bf B601} (2001) 248--286
  [\href{http://arXiv.org/abs/hep-ph/0008287}{{\tt hep-ph/0008287}}].

\bibitem{Gehrmann:2001ck}
T.~Gehrmann and E.~Remiddi, {\it {Two loop master integrals for gamma* --> 3
  jets: The Nonplanar topologies}},  {\em Nucl. Phys.} {\bf B601} (2001)
  287--317 [\href{http://arXiv.org/abs/hep-ph/0101124}{{\tt hep-ph/0101124}}].

\bibitem{Duhr:2014woa}
C.~Duhr, {\it {Mathematical aspects of scattering amplitudes}},  in {\em
  {Proceedings, Theoretical Advanced Study Institute in Elementary Particle
  Physics: Journeys Through the Precision Frontier: Amplitudes for Colliders
  (TASI 2014): Boulder, Colorado, June 2-27, 2014}}, pp.~419--476, 2015.
\newblock \href{http://arXiv.org/abs/1411.7538}{{\tt 1411.7538}}.

\bibitem{Catani:1998bh}
S.~Catani, {\it {The Singular behavior of QCD amplitudes at two loop order}},
  {\em Phys. Lett.} {\bf B427} (1998) 161--171
  [\href{http://arXiv.org/abs/hep-ph/9802439}{{\tt hep-ph/9802439}}].

\bibitem{Li:2013lsa}
Y.~Li and H.~X. Zhu, {\it {Single soft gluon emission at two loops}},  {\em
  JHEP} {\bf 11} (2013) 080 [\href{http://arXiv.org/abs/1309.4391}{{\tt
  1309.4391}}].

\bibitem{Badger:2004uk}
S.~D. Badger and E.~W.~N. Glover, {\it {Two loop splitting functions in QCD}},
  {\em JHEP} {\bf 07} (2004) 040
  [\href{http://arXiv.org/abs/hep-ph/0405236}{{\tt hep-ph/0405236}}].

\bibitem{Duhr:2014nda}
C.~Duhr, T.~Gehrmann and M.~Jaquier, {\it {Two-loop splitting amplitudes and
  the single-real contribution to inclusive Higgs production at N$^3$LO}},
  {\em JHEP} {\bf 02} (2015) 077 [\href{http://arXiv.org/abs/1411.3587}{{\tt
  1411.3587}}].

\bibitem{Campbell:1999ah}
J.~M. Campbell and R.~K. Ellis, {\it {An Update on vector boson pair production
  at hadron colliders}},  {\em Phys. Rev.} {\bf D60} (1999) 113006
  [\href{http://arXiv.org/abs/hep-ph/9905386}{{\tt hep-ph/9905386}}].

\bibitem{Campbell:2011bn}
J.~M. Campbell, R.~K. Ellis and C.~Williams, {\it {Vector boson pair production
  at the LHC}},  {\em JHEP} {\bf 07} (2011) 018
  [\href{http://arXiv.org/abs/1105.0020}{{\tt 1105.0020}}].

\bibitem{Campbell:2015qma}
J.~M. Campbell, R.~K. Ellis and W.~T. Giele, {\it {A Multi-Threaded Version of
  MCFM}},  {\em Eur. Phys. J.} {\bf C75} (2015), no.~6 246
  [\href{http://arXiv.org/abs/1503.06182}{{\tt 1503.06182}}].

\bibitem{Boughezal:2016wmq}
R.~Boughezal, J.~M. Campbell, R.~K. Ellis, C.~Focke, W.~Giele, X.~Liu,
  F.~Petriello and C.~Williams, {\it {Color singlet production at NNLO in
  MCFM}},  {\em Eur. Phys. J.} {\bf C77} (2017), no.~1 7
  [\href{http://arXiv.org/abs/1605.08011}{{\tt 1605.08011}}].

\bibitem{GehrmannDeRidder:2008ug}
A.~Gehrmann-De~Ridder, T.~Gehrmann, E.~W.~N. Glover and G.~Heinrich, {\it {Jet
  rates in electron-positron annihilation at O(alpha(s)**3) in QCD}},  {\em
  Phys. Rev. Lett.} {\bf 100} (2008) 172001
  [\href{http://arXiv.org/abs/0802.0813}{{\tt 0802.0813}}].

\bibitem{Weinzierl:2008iv}
S.~Weinzierl, {\it {NNLO corrections to 3-jet observables in electron-positron
  annihilation}},  {\em Phys. Rev. Lett.} {\bf 101} (2008) 162001
  [\href{http://arXiv.org/abs/0807.3241}{{\tt 0807.3241}}].

\bibitem{Catani:2000pi}
S.~Catani and M.~Grazzini, {\it {The soft gluon current at one loop order}},
  {\em Nucl. Phys.} {\bf B591} (2000) 435--454
  [\href{http://arXiv.org/abs/hep-ph/0007142}{{\tt hep-ph/0007142}}].

\end{thebibliography}\endgroup

\end{document}